\documentclass[aps,prb,twocolumn,superscriptaddress,10pt,showpacs]{revtex4-1}

\usepackage[utf8x]{inputenc}
\usepackage{graphicx}
\usepackage{amsmath,amsthm,amssymb}
\usepackage[dvipsnames]{xcolor}
\usepackage[colorlinks,linkcolor=blue,citecolor=blue,urlcolor=blue]{hyperref}
\usepackage{physics}

\usepackage{subfigure}

\graphicspath{{./../images/}}

\newcommand{\bfs}{\boldsymbol}
\newcommand{\ha}{\hat{a}}
\newcommand{\hA}{\hat{A}}
\newcommand{\hc}{\hat{c}}

\newcommand{\hH}{\hat{H}}
\newcommand{\hh}{\hat{h}}

\newcommand{\OP}[1]{\skew{4}\hat{#1}}

\newcommand{\mc}[1]{\mathcal{#1}}

\definecolor{darkpink}{rgb}{0.91, 0.33, 0.5}

\begin{document}

\title{Non-linear optical processes in cavity light-matter systems}
\author{Markus Lysne}
\affiliation{Department of Physics, University of Fribourg, 1700 Fribourg, Switzerland}
\author{Philipp Werner}
\affiliation{Department of Physics, University of Fribourg, 1700 Fribourg, Switzerland}

\date{\today}


\begin{abstract}
We study non-linear optical effects in  
electron systems 
with and without inversion symmetry in a Fabry-Perot cavity. General photon up- and down-conversion processes are modeled by the coupling of a noninteracting lattice model to two  
modes of the quantized light field. 
Effective descriptions retaining the most relevant states are devised via  
downfolding 
and 
a generalized Householder transformation. 
These models are used to relate the transition 
amplitudes 
for even order 
photon-conversion processes to the shift vector, a topological quantity describing the difference in polarization between the valence and conduction band in non-centrosymmetric systems. 
We also demonstrate that the truncated models, despite their small Hilbert space, capture correlation effects induced by the photons in the electronic subsystem.

\end{abstract}

\maketitle

\section{Introduction}
\label{sec:introduction}
Understanding the properties of light-matter coupled systems is a long-standing challenge in condensed matter physics. 
The goal of controlling material properties with classical light has been pursued actively in recent years, and remarkable phenomena such as interaction tuning or qualitative changes in the band structure via Floquet engineering have been theoretically demonstrated.\cite{Oka_2009, Lindner_2011, Tsuji_2011, Bukov_2015, Mentink_2015, Claassen_2017, Dasari_2018, Oka_2019} On the spectroscopy side, research on high-harmonic generation (HHG) has shown that non-linear optical processes inherit important fingerprints of the electronic structure 
and even the Berry curvature of solids.\cite{Ghimire_2011, Luu_2015, Vampa_2015b, Silva_2018, Murakami_2018, Luu_2018}

Another stimulating prospect is the exploration and control of strongly coupled light-matter systems in cavities, where the quantized nature of the photon field plays an important role.\cite{Basov_2016, Dufferwiel_2015, Mak_2016, Liu_2015, Sentef_2020}  
Despite decades of studies of cavity quantum electrodynamics (CQED) problems, as exemplified by the Rabi or Dicke model,\cite{Garraway_2011} this field has attracted renewed interest in the condensed matter community with the discussion of purported super-radiant states and the possibility of engineering novel states of matter.\cite{Georges_2019, Li_2020, Ashida_2020, Sentef_2018, Latini_2021} 
With phenomena such as HHG in mind, one may ask how photon up- and down-conversion occurs in these fully quantized light-matter systems, and what these nonlinear phenomena reveal about the (topological) properties of the material. 

Following up on the pioneering work of Sipe {\it et al.},\cite{Sipe_2000, Aversa_1995} 
a recent Floquet study by Morimoto and Nagaosa \cite{Morimoto_2016} showed that in systems with time-reversal symmetry (TRS), but without inversion symmetry (IS), nonlinear optical processes can be related to topological quantities.  Specifically, it was demonstrated that Floquet theory for an effective two-band model provides a suitable framework to investigate the shift current and non-linear Hall conductivity.  
Here, we extend this approach to a cavity set-up with quantized light. Instead of Floquet sidebands, we will consider a low energy theory in a system with two dominant photon modes. 
This simple set-up allows us to derive expressions for photon up- and down-conversion processes which are analogous to those in Refs.~\onlinecite{Morimoto_2016, Sipe_2000, Aversa_1995}. 

We further show that it is possible to capture the most relevant effects of the photons on the electronic states in an effective description involving a small number of ``molecular orbitals." This description is obtained by a block Householder transformation, which enables systematic truncations of the Hilbert space. Even after a truncation to just four states, the effective model correctly describes the photon-conversion processes, and provides qualitatively correct results for the kinetic energy and charge correlation functions. 

The paper is organized as follows. In Sec.~\ref{sec:model} we derive our minimal model for photon conversion in solids interacting with quantized light. This model is downfolded to an effective photon model in Sec.~\ref{sec:photConv}, and it is shown that the transition amplitudes for even order up- or down-conversion are related to the shift vector. In Sec.~\ref{sec:eff_el} we introduce the block Householder transformation and derive few-states electron-photon models. Section~\ref{sec:results} tests the few-states effective description against the full model for a one-dimensional chain coupled to two photon modes.

\section{Model} 
\label{sec:model}

\subsection{Coulomb gauge Hamiltonian}

We consider a matter Hamiltonian representing a noninteracting lattice model with two orbitals in each unit cell. 
To describe the coupling to an electromagnetic field, we employ here the Coulomb gauge\cite{Stokes_2020} (similar to the velocity gauge in Ref.~\onlinecite{Ventura_2017}). The form of our Hamiltonian is thus different from the Hamiltonians encountered in recent studies which employ either a dipole gauge, obtainable through a Power-Zienau-Wolley (PZW) \cite{Li_2020, Schuler_2021} gauge transformation of the Coulomb gauge Hamiltonian, or the multi-center PZW transformation which preserves translational invariance.\cite{Golez_2019, Li_2020, Schuler_2021} We refer to Appendix~\ref{sec:appendixGauge} for a discussion of the 
mappings between these different representations.

The Coulomb gauge Hamiltonian in a second quantized form can be written as\cite{Ventura_2017}
\begin{align}
	\hat{H}_{CG} =& \sum_{k,\alpha,\beta} \hc_{k,\alpha}^\dagger \big\langle u_{k,\alpha} \big| \hh_0\big(k-q\textstyle{\sum_\mu} g_\mu \hA_\mu \big) \big| u_{k,\beta}\big\rangle \hc_{k,\beta} \nonumber\\
	&+ \sum_{\mu} \frac{\Omega_\mu}{2} \big( \hat{\Pi}_\mu^2 + \hA_\mu^2 \big), 
	\label{eq:fullHamCG}
\end{align}
with $\mu$ denoting the different modes of the transverse electromagnetic field. The vector potential and its conjugate variable are defined as $\hA_\mu = \tfrac{1}{\sqrt{2}}(\ha_\mu+\ha_\mu^\dagger)$ and $\hat{\Pi}_\mu = \tfrac{i}{\sqrt{2}}(\ha_\mu^\dagger-\ha_\mu)$ in terms of the photon creation (annihilation) operators $a^\dagger_\mu$ ($a_\mu$), 
and satisfy the commutation relations $[\hA_\nu, \hat{\Pi}_\mu]=i\delta_{\mu,\nu}$ (all other commutators being zero).
 The photon energy for mode $\mu$ is $\Omega_\mu$, and the photon coupling (which we assume to be real) is denoted by $g_\mu$. 
$\hc_{k,\alpha}^\dagger$ creates an electron in a Bloch state with momentum $k$ and band index $\alpha$: $\hc_{k,\alpha}^\dagger\ket{\text{vac}}=\ket{\psi_{k,\alpha}}=e^{ik\hat r}\ket{u_{k,\alpha}}$, 
$\hh_0(k)\ket{u_{k,\alpha}} =\epsilon_{\alpha}(k)\ket{u_{k,\alpha}}$, with 
$\epsilon_{\alpha}(k)$ the corresponding energy. Here, 
$\hat h_0(k)=e^{-ik\hat r}\hat H_0 e^{ik\hat r}$ 
with $\hat H_0$ the Hamiltonian of the noninteracting matter system.  
We will use units where the charge $q=-\abs{e}=-1$.

\begin{figure}[t]
\centering
\includegraphics[width=0.8\columnwidth]{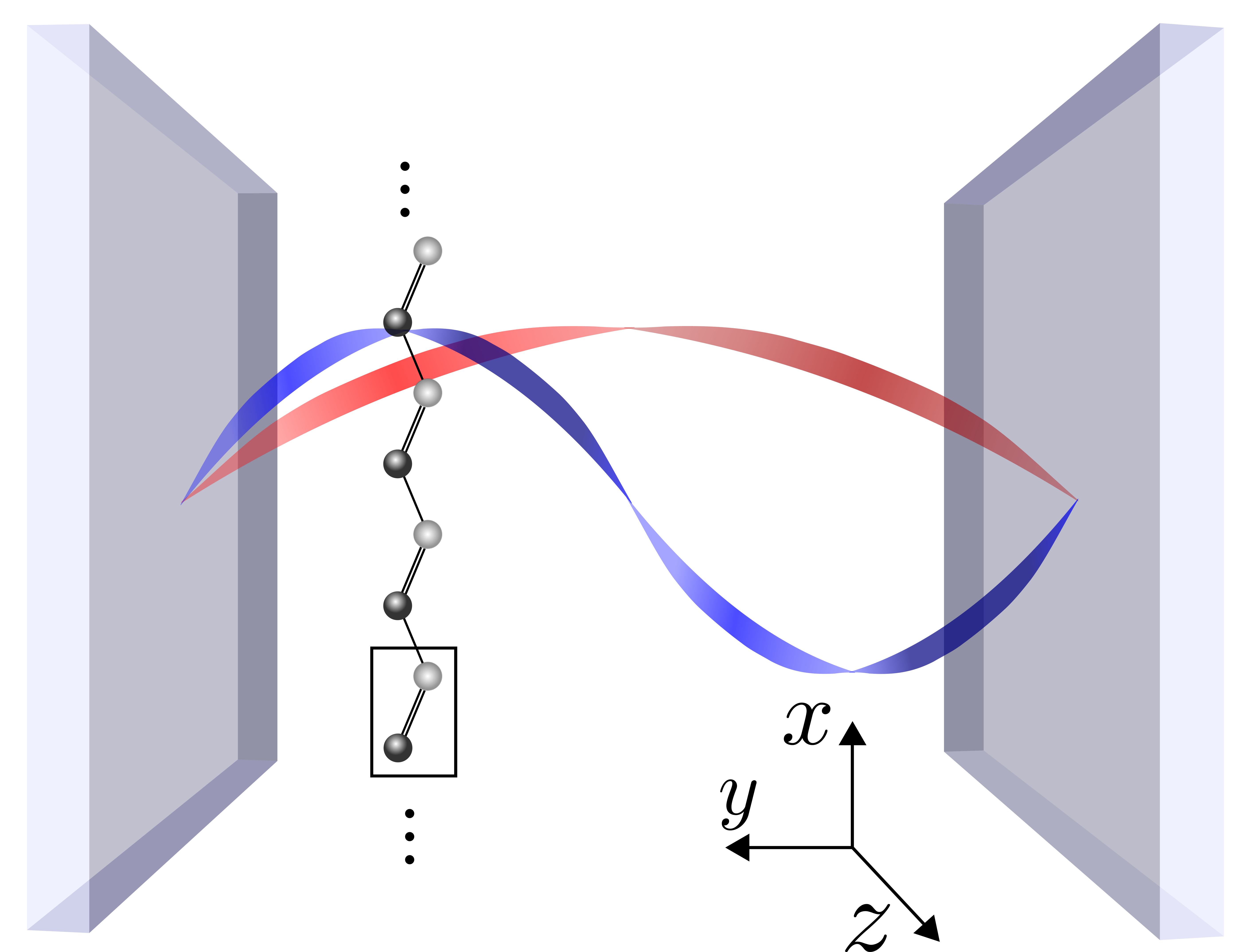} \\
\vspace{5mm}
\includegraphics[width=0.9\columnwidth]{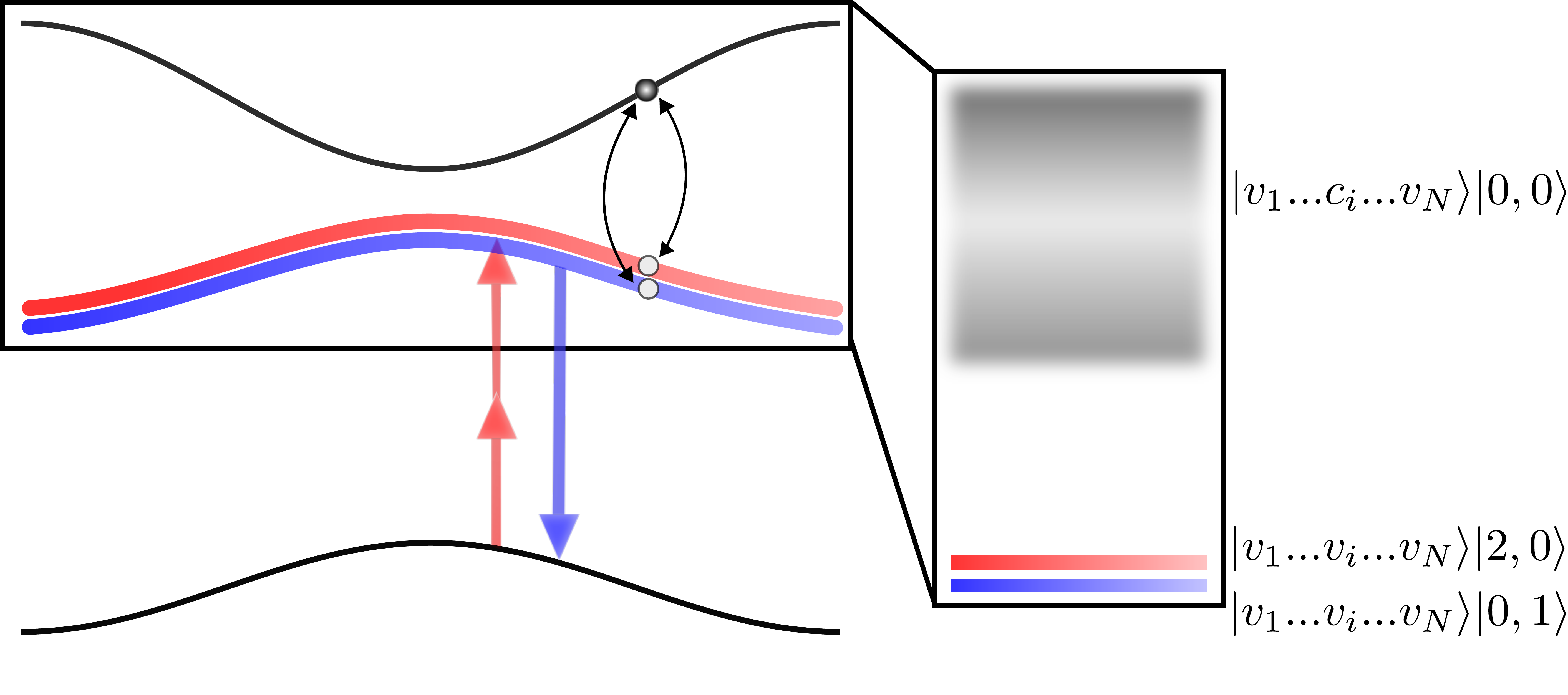}
\caption{Top panel: Sketch of a Fabry Perot cavity with two relevant photon modes, and a one-dimensional material placed at an antinode of the second mode. The box indicates a possible choice of unit cell. Bottom panel: Illustration of the undressed valence and conduction bands (black lines), and the valence band dressed with two additional photons of energy $\Omega_1$ (red), as well as the valence band dressed with one photon of energy $\Omega_2$ (blue). Here, the box encloses the states considered in the low-energy model. 
The bottom right part of the figure illustrates the energy levels of the many-body system (see text).
}   
\label{fig:bandStructure}
\end{figure}

For the study of general nonlinear optical processes involving two modes (see top panel of Fig.~\ref{fig:bandStructure}), one may expand 
\begin{align}
	&\hat{h}_0(k  + g_1\hat A_1  + g_2\hat A_2) \nonumber\\
	&\quad= \sum_{m,n} \frac{1}{m! n!} \frac{\partial^{m+n} \hat{h}_0(k)}{\partial k^{m+n}} (g_1\hat{A}_1)^{m} (g_2\hat{A}_2)^{n},\label{eq:genExp}
\end{align}
were we  
assume that 
each term for $n$ or $m>0$ in Eq.~\eqref{eq:genExp} scales as $\tfrac{1}{\sqrt{N}}$ with lattice size $N$.
Specifically, we will consider a two-band system with a conduction band ($\alpha=c$) and a valence band ($\alpha=v$), and we will be primarily interested in photon up- and down-conversion processes where $\Omega_2=2\Omega_1 
<\text{min}_k(\epsilon_{k,c}- \epsilon_{k,v})$ (splitting between the conduction and valence band, see bottom left panel of Fig.~\ref{fig:bandStructure}), and where the couplings $g_\mu$ are so small that we only need to consider states with $0$ or $2$ ($1$) photons in mode $\mu=1$ ($\mu=2$). In this case, one can restrict the sum in Eq.~\eqref{eq:genExp} to the first order term in $\hat A_2$ and the second order term in $\hat A_1$ and write
\begin{align}
	&\hat{h}_0(k + g_1\hat{A}_1 + g_2\hat{A}_2) 
	\approx \hat h_0(k) + \hat{v}(k) g_2\hat{A}_2 + \tfrac{1}{2} \hat{v}'(k) g_1^2 \hat{A}_1^{2}  \label{eq:pertStart}
\end{align}
with $\hat{v}(k)=\partial_k \hat{h}_0(k)$ and $\hat v'(k)=\partial^2_k \hat{h}_0(k)$. 
Note that the latter two operators may have off-diagonal elements in the band basis, since generically $\hH_0$ does not commute with $[\hat r,\hH_0]$ and $[\hat{r},[\hat r,\hH_0]]$, so that there is no complete set of common eigenstates of $\hat h_0(k)$ and $\hat{v}(k)$ or $\hat h_0(k)$ and $\hat v'(k)$. 

In addition to the weak coupling assumption, we will employ the rotating wave approximation (RWA) and neglect terms such as $\hc_{k,v}^\dagger \hc_{k,c}g_2\ha_2$,  $\hc_{k,v}^\dagger \hc_{k,c}g_2\ha_1^2$, etc. 
Also the contribution from $\ha_1^\dagger \ha_1$ in the expansion of $\hat{A}_1^2$, which leads to a level renormalization $O(g_1^2)$, will be neglected. With these approximations, the Coulomb gauge Hamiltonian becomes 
\begin{equation} \label{eq:pertStartRWA}
\begin{aligned}
	&\hat{H}_{CG} \approx \sum_k \hat{n}_{k,\alpha} \epsilon_{\alpha}(k) + \sum_k \Big[ \frac{g_2}{\sqrt{2} }v_{vc}(k) \hc_{k,v}^\dagger \hc_{k,c}\ha_2^\dagger \\
	&\quad+ \frac{g_1^2}{4}  v_{vc}^\prime (k) \hc_{k,v}^\dagger \hc_{k,c}  (\ha_1^\dagger)^2 + \mathrm{h.c.}\Big] + \sum_\mu \frac{\Omega_\mu}{2} \big( \hat{\Pi}_\mu^2 + \hA_\mu^2 \big), \\
\end{aligned}
\end{equation}
with $v_{\alpha\beta}(k)=\langle u_{k,\alpha}|\hat v(k)|u_{k,\beta}\rangle$ and similarly for $v'_{\alpha\beta}(k)$. 
The photon coupling strengths $g_2$ and $g_1$ are assumed to scale with system size $N$ as $g_2 =\tilde g_2/\sqrt{N}$ and $g_1^2 =\tilde g_1^2/\sqrt{N}$ where $\tilde{g_1}$ and $\tilde{g_2}$ are fixed parameters.

The specific set-up which we have in mind is illustrated in Fig.~\ref{fig:bandStructure} and consists of a one-dimensional material placed within a Fabry Perot cavity with two relevant photon modes. The polarization of the modes is along the direction of the chain ($x$ axis) and the propagation direction perpendicular to the chain ($y$ axis). By placing the sample at an antinode of the higher energy mode,  
we can assure a coupling to both modes. 
Additionally, the dipole approximation is employed along the direction of the chain. In this situation, the canonical commutation relations read $[\hat{A}_x(y), \hat{\Pi}_x(y’)]=i\delta(y-y’)$. \cite{Lenk_2020} A convenient mode expansion is $\hat{A}_x(y) = \sum_\mu \phi_\mu(y)\hat{A}_\mu$, $\hat{\Pi}_x(y) = \sum_\mu \epsilon(y) \phi_\mu^*(y) \hat{\Pi}_\mu$ with $\epsilon(y)$ the dielectric function in the cavity, $[\hat{A}_\mu, \hat{\Pi}_\nu]=i\delta_{\mu,\nu}$, and $\int dy \epsilon(y) \phi_{\mu}^*(y) \phi_\nu(y) = \delta_{\mu,\nu}.$ \cite{Li_2020, Glauber_1991} As noted in Ref.~\onlinecite{Glauber_1991}, the free-field part of Eq.~\eqref{eq:fullHamCG} will depend on $\epsilon(y)$ in dielectric media. In the following, we will use $\epsilon=1$ for simplicity. If $\epsilon(y)\neq 1$, one can apply a suitable scaling transformation on the variables in Eq.~\eqref{eq:fullHamCG}. The cavity may be finite or infinitely extended along the directions $x$ and $z$, as long as there is no mixing with any modes with wavevectors $q_{x,z}\neq 0$.

\subsection{Low energy model}

For the electron-photon coupled many-body system, it is convenient to 
introduce the basis states $\ket{\alpha}\ket{m,n}$, where $\ket{\alpha} \equiv \otimes_{i=1}^N \ket{\alpha_i}$ with $\ket{\alpha_i} \in \{|c_i\rangle\equiv|\psi_{k_i,c}\rangle,|v_i\rangle\equiv|\psi_{k_i,v}\rangle\}$ representing an electron with momentum $k_i$ in either the conduction or valence band, and $m$ ($n$) the number of photons with frequency $\Omega_1$ ($\Omega_2$). The relevant states for photon up- and down-conversion can be identified by looking at the bottom panel in Fig.~\ref{fig:bandStructure}. This figure 
displays the band structure, including photon-dressed states, and suggests to consider a 
subspace (black box) comprised of conduction band states (states of the type $\ket{v_1,\ldots,v_{i-1},c_i,v_{i+1},\ldots,v_N}\ket{0,0}$) as well as the valence band dressed with $1$ or $2$ photons of frequency $\Omega_2$ and $\Omega_1$, respectively (states $\ket{v_1, \ldots,v_N}\ket{0,1}$ and $\ket{v_1,\ldots,v_N}\ket{2,0}$). 
These many-body states, along with their energies, are sketched in the bottom right panel of the figure. 
The corresponding low-energy model is described by the Hamiltonian matrix $H^\text{low}$ 
 expressed in the basis $\{ \ket{v_1, v_2, ..., v_N}\ket{0,1}$$,$ $\ket{v_1, v_2, ..., v_N}\ket{2,0}$$,$ $\{\ket{v_1, v_2, ...,c_i, ..., v_N}\ket{0,0} \} \}$, which reads
\begin{widetext}
\begin{equation} \label{eq:fullHam}
{H}^\text{low}=
\begin{pmatrix}
	\Omega_2 & 0 & \tfrac{{{g}}_2}{\sqrt{2}} v_{vc}(k_1) & \tfrac{{{g}}_2}{\sqrt{2}} v_{vc}(k_2) & \dots & \tfrac{{{g}}_2}{\sqrt{2}} v_{vc}(k_N) \\
	0 & 2\Omega_1 &\tfrac{{{g}}_1^2}{2\sqrt{2}} {v}^\prime_{vc}(k_1) & \tfrac{{{g}}_1^2}{2\sqrt{2}} v_{vc}^{\prime}(k_2) & \dots & \tfrac{{{g}}_1^2}{2\sqrt{2}} v_{vc}^{\prime}(k_N) \\
	\tfrac{{{g}}_2}{\sqrt{2}} v_{cv}(k_1) & \tfrac{{{g}}_1^2}{2\sqrt{2}} v_{cv}^{\prime}(k_1) & \epsilon_c(k_1) - \epsilon_v(k_1) & 0 & \dots &0\\
	\tfrac{{{g}}_2}{\sqrt{2}} v_{cv}(k_2) & \tfrac{{{g}}_1^2}{2\sqrt{2}} v_{cv}^{\prime}(k_2) &0 & \epsilon_c(k_2) - \epsilon_v(k_2) &  &  \\
	\vdots & \vdots & \vdots & & \ddots & \vdots \\
	\tfrac{{{g}}_2}{\sqrt{2}} v_{cv}(k_N) & \tfrac{{{g}}_1^2}{2\sqrt{2}} v_{cv}^{\prime}(k_N) & 0 & & \dots & \epsilon_c(k_N) - \epsilon_v(k_N) \\
\end{pmatrix}.	
\end{equation}		
\end{widetext}

\section{Photon conversion and shift vector}
\label{sec:photConv}

Even though model (\ref{eq:fullHam}) has no direct coupling between the different photon states, photon conversion processes are induced via the coupling to electron-hole excitations. 
Similar phenomena are observed in HHG experiments on semi-conductors,\cite{Ghimire_2019} where electron excitation and de-excitation processes result in photon up-conversion. 
More specifically, the conversion from two $\Omega_1$ photons to a single $\Omega_2$ photon in our cavity set-up is reminiscent of second harmonic generation in semiconductors (without inversion symmetry), where electrons are excited from the valence to the conduction band by an electric field with frequency $\Omega_1$, while the de-excitation process generates radiation with frequency $\Omega_2=2\Omega_1$. 
Analogous photon up- and down-conversions have also been discussed in atomic physics.\cite{Kockum_2017} 

Following Ref.~\onlinecite{Kockum_2017}, we introduce the states $\ket{i}, \ket{f}$ representing the initial and final states of the photon conversion process, respectively, and evaluate the transition amplitude through time-dependent perturbation theory. This calculation is valid if $\ket{i}, \ket{f}$ are similar in energy and their energy difference to other states $|j\rangle$ is sufficiently large. Splitting the Hamiltonian in Eq.~\eqref{eq:pertStartRWA} into 
\begin{equation}
	\hat{H}_1 = \Omega_1 \hat{a}_1^\dagger \hat{a}_1+\Omega_2 \hat{a}_2^\dagger \hat{a}_2 + \sum_{k,\alpha} \epsilon_{\alpha}(k) \hat{n}_{k,\alpha},
\end{equation}
where we have used that $\frac{\Omega_\mu}{2} \big( \hat{\Pi}_\mu^2 + \hA_\mu^2 \big)=\Omega_\mu \ha_\mu^\dagger \ha_\mu$ up to a constant, and 
\begin{align}
	\hat{H}_{2} =& \frac{g_2\hat{a}_2}{\sqrt{2}}\sum_k v_{cv} (k) \hat{c}_{k,c}^\dagger \hat{c}_{k,v}  \nonumber\\
	&
	+\frac{g_1^2\hat{a}_1^2}{4} \sum_k v_{cv}^{\prime} (k) \hat{c}_{k,c}^\dagger \hat{c}_{k,v}+  \text{h.c.},
\end{align}
we obtain in second order perturbation theory the effective photon model
\begin{equation}
	\hat H_\text{ph} = d_\text{trans} {\ket{f}\!\bra{i}}+ d_\text{trans}^*{\ket{i}\!\bra{f}}+d_z({\ket{f}\!\bra{f}} - {\ket{i}
	\!\bra{i}} ), 
\end{equation}
with 
\begin{equation} \label{eq:geffExample}
	d_\text{trans} = \sum_{j\neq i,f} \frac{\bra{f}H_2 \ket{j}\bra{j} H_2 \ket{i} }{E_i-E_j }.
\end{equation}
Equation~\eqref{eq:geffExample} corresponds to the amplitude for second order photon conversion if we choose
$|i\rangle=\ket{2,0}$ and $|f\rangle=\ket{0,1}$, where $|m,n\rangle\equiv |v_1,\ldots,v_N\rangle|m,n\rangle$, and $E_i=2\Omega_1+E_v=\Omega_2+E_v=E_f$ ($E_v$ denotes the energy of the full valence band). 
The state $|j\rangle$ describes the system without photons, but the electron with momentum $k_j$ in the conduction band, which has energy $E_j=\epsilon_c(k_j)-\epsilon_v(k_j)+E_v$. Hence 
$\bra{j}\hat{H}_{2}\ket{2,0} = \frac{{g}_1^2}{2\sqrt{2}}(v')_{cv} (k_j)$, 
$\bra{0,1}\hat{H}_2\ket{j} = \tfrac{g_2}{\sqrt{2}} v_{vc} (k_j),$
and we get 
\begin{equation} \label{eq:geff}
\begin{aligned}
	d^{(2)}_\text{trans} 
	=\frac{{g}_1^2 {g}_2}{4} \sum_k \frac{v_{vc}(k) (v')_{cv}(k) }{2\Omega_1 -(\epsilon_c(k)-\epsilon_v(k))}, \\
\end{aligned}
\end{equation}
where the superscript `(2)' indicates that this is the amplitude for the second order photon conversion process ($\Omega_2=2\Omega_1$). Furthermore, $d_z^{(2)}$ is given by
\begin{equation}
\begin{aligned}
	d_z^{(2)} 
	&=\frac{1}{2}\sum_k \frac{ ( \tfrac{g_2}{\sqrt{2}} \abs{v_{vc}(k)})^2 - ( \tfrac{g_1^2}{2\sqrt{2}}\abs{(v')_{vc}(k)})^2 }{2\Omega_1 - (\epsilon_c(k) - \epsilon_v(k))}. \\
\end{aligned}
\end{equation}

Assuming time reversal symmetry ($v_{\alpha\alpha}(-k) =-v_{\alpha\alpha}(k)$ and $\epsilon_{\alpha}(-k) =\epsilon_{\alpha}(k)$) one can show that $d^{(2)}_\text{trans}$ is purely imaginary. In fact, in the numerator of the sum in Eq.~(\ref{eq:geff}) we recognize the product\cite{Morimoto_2016} 
\begin{align}
	&v_{vc}(k) (v')_{cv}(k)  \\
	&= \abs{v_{vc}}^2 \Big[\partial_k \mathrm{log}v_{cv}- i(\xi_c-\xi_v) + \frac{v_{cc}-v_{vv}}{\epsilon_c-\epsilon_v}\Big], \label{eq_product}
\end{align}
where we have used that
	$(v')_{cv}=\big( \frac{\partial v}{\partial k}\big)_{cv} = \partial_k v_{cv} 
	\quad - \bra{\partial_k u_{k,c}}\hat{v}(k)\ket{u_{k,v}} - \bra{u_{k,c}}\hat{v}(k)\ket{\partial_k u_{k,v}} $
and inserted $\ket{u_{k,v}}\bra{u_{k,v}} + \ket{u_{k,c}}\bra{u_{k,c}}=1$ in order to obtain the last two terms above. Furthermore, we note that
\begin{equation}
\bra{u_{k,c}} \partial_k \ket{u_{k,v}} = - \frac{v_{cv}(k)}{\epsilon_c(k)-\epsilon_v(k)}
\end{equation}
and recall the definition of the Berry connection \cite{Xiao_2010, Ventura_2017} 
\begin{equation}
\xi_\alpha(k) = i\bra{u_{k,\alpha}}\partial_k \ket{u_{k,\alpha}},
\end{equation} 
which is real, since $\partial_k \bra{u_{k,\alpha}}\ket{u_{k,\alpha}} = \bra{\partial_k u_{k,\alpha}}\ket{u_{k,\alpha}} +  \bra{u_{k,\alpha}}\ket{\partial_k u_{k,\alpha}}=0$.
Using that $\text{Re}[\partial_k \log v_{cv}]=\text{Re}[\partial_k |v_{vc}|/|v_{vc}|]$, that $\partial_k |v_{vc}|$ ($|v_{vc}|$) is an odd (even) function of $k$, and that the denominator is real and even, one thus finds that the real parts in the sum cancel. The remaining imaginary terms can be expressed as
\begin{align}
	d^{(2)}_\text{trans} 
	= i \frac{{g}_1^2 {g}_2}{4} \sum_k \frac{\abs{v_{vc}(k)}^2 R_k^{cv} }{2\Omega_1 - (\epsilon_c(k) - \epsilon_v(k))}. \label{g2_simplified}
\end{align}
In Eq.~(\ref{g2_simplified}) we introduced the shift vector $R_k^{cv}$, defined as \cite{Morimoto_2016} 
\begin{equation}
	R_k^{cv} = \partial_k \textrm{Im}(\log v_{cv}) -(\xi_c - \xi_v).
\end{equation}
The shift vector is a quantity which appears in the second order optical response of systems that break inversion symmetry and its integral over the Brillouin zone can be related to the polarization difference between the valence and conduction band.\cite{Fregoso_2017}

\begin{figure}[t]
\centering
\includegraphics[width=0.8\columnwidth]{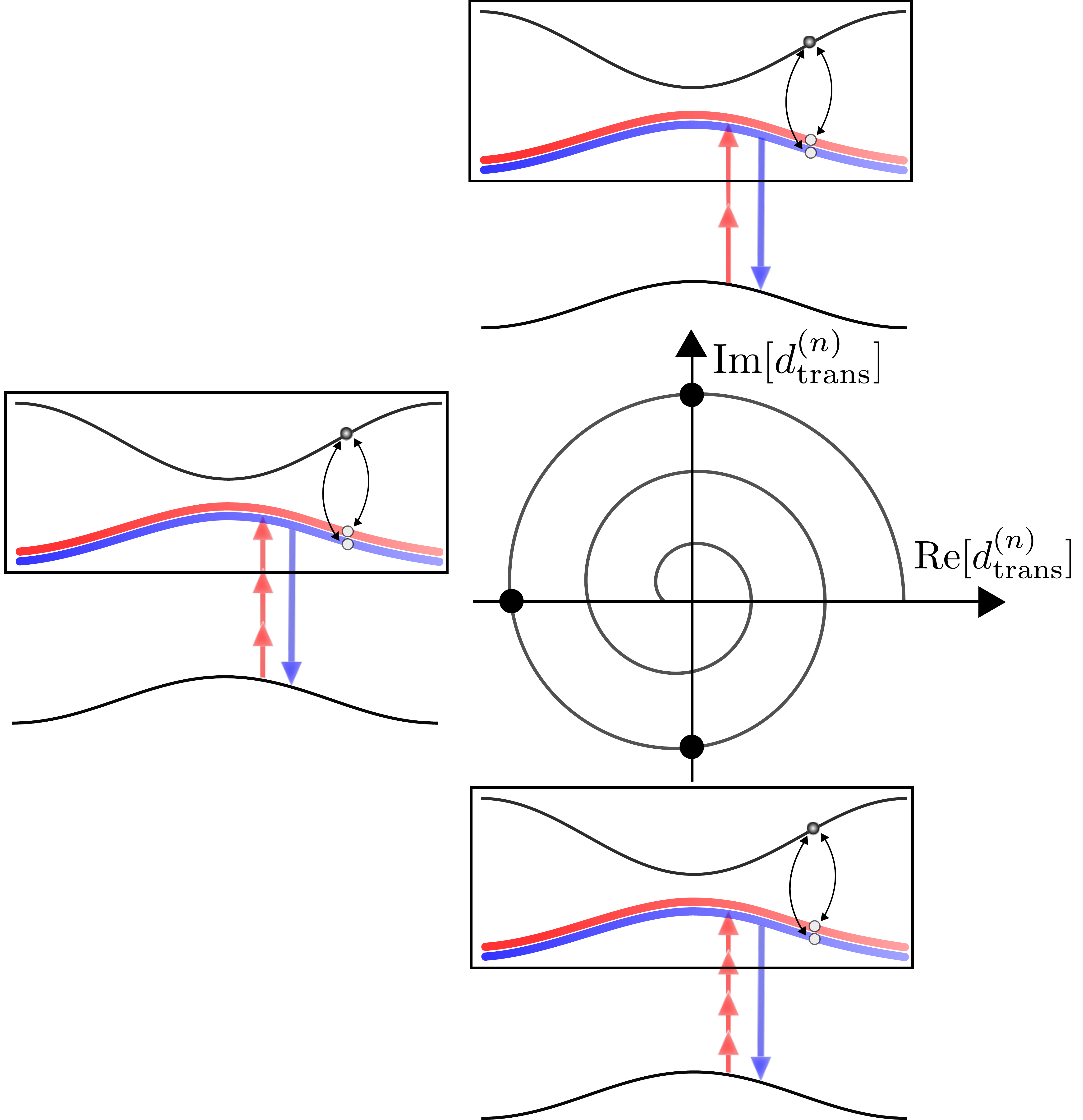}
\caption{Depiction of different low energy models corresponding to second, third and fourth order photon conversion, and the corresponding values of  
$\boldsymbol{d}$ projected onto the $x$-$y$ plane. 
}   
\label{fig:snail}
\end{figure}

These results can be readily generalized to higher order photon conversion processes, although the low energy models derived from them may be more difficult to justify. If the mode frequencies are tuned to satisfy the resonance condition $\Omega_2 =n\Omega_1$, with $n>1$ some integer, the low energy model is given by
\begin{equation} \label{eq:pertStartRWA_n}
\begin{aligned}
	\hat{H}_{CG} \approx& \sum_k \hat{n}_{k,\alpha} \epsilon_{\alpha}(k) + \sum_k \Bigg[ \frac{g_2}{\sqrt{2}}v_{vc}(k) \hc_{k,v}^\dagger \hc_{k,c}\ha_2^\dagger \\
	&+ \frac{g_1^n}{{2}^\frac{n}{2} n!} \Big(\frac{\partial^n \hat{h}(k) }{\partial k^n}\Big)_{vc} \hc_{k,v}^\dagger \hc_{k,c} (\ha_1^\dagger)^n + \mathrm{h.c.}\Bigg] \nonumber\\
	&+ \sum_\mu \frac{\Omega_\mu}{2} \big( \hat{\Pi}_\mu^2 + \hA_\mu^2 \big), \\
\end{aligned}
\end{equation}
and $\hat H_\text{ph}$ involves the coupling
\begin{equation} \label{eq:geffGen}
\begin{aligned}
	d^{(n)}_\text{trans} 
	=\frac{{g}_1^n {g}_2}{2^\frac{n+1}{2}\sqrt{n!}} \sum_k \frac{v_{vc}(k)(v^{(n-1)})_{cv}(k) }{n\Omega_1 -(\epsilon_c(k)-\epsilon_v(k))}, \\
\end{aligned}
\end{equation}
where we introduced $\hat v^{(n-1)}(k)=\frac{\partial^n \hat{h}(k)}{\partial k^n}$. 
If $\hat{h}(k)$ is time-reversal symmetric and comprised of trigonometric functions, then $\partial_k^3 \hat{v}(k) = -\partial_k \hat{v}(k)$, which leads to 
\begin{align}
	&v_{vc}(v^{(n-1)})_{cv} 
	=  \Big[\sin(\frac{n\pi}{2}) \abs{v_{vc}}^2 - \cos(\frac{n\pi}{2}) v_{vc}(v')_{cv} \Big]. \label{eq:genPattern}
\end{align}
The calculation 
of $d^{(n)}_\text{trans}$ and $d_z^{(n)}$ then proceeds as before. In particular, if we repeat the calculation for third order processes ($n=3$), assuming $\Omega_2=3\Omega_1$, we find that $d^{(3)}_\text{trans}$ is real, and not related to the shift vector:
\begin{equation} \label{eq:geff3}
	d^{(3)}_\text{trans}=-\frac{{g}_1^3 {g}_2}{4\cdot\sqrt{3!}} \sum_k \frac{ \abs{v_{vc}(k)}^2 }{3\Omega_1 -(\epsilon_c(k)-\epsilon_v(k))}. 
\end{equation}
For even $n$, the result is purely imaginary, and nonzero only in the presence of a non-vanishing shift vector, while for $n$ odd, the result is real and independent of the shift vector. As illustrated in Fig.~\ref{fig:snail}, we can define a vector $\boldsymbol{d}=(d_x,d_y,d_z)$ with $d_x = \text{Re}[d^{(n)}_\text{trans}]$, $d_y = \text{Im}[d^{(n)}_\text{trans}]$, $d_z=d_z^{(n)}$, which rotates (in steps of 90$^\circ$) around the $z$-axis in a  generically elliptic spiral pattern as $n$ increases. If $d_z$ is nonzero, as is usually the case, the expectation values of $\sigma_x$, $\sigma_y$ in the downfolded model are less than one. 

In Appendix~\ref{app_downfolding} we show that the same effective photon model $\hat H_\text{ph}$ can also be obtained from the low-energy model (\ref{eq:fullHam}) by a downfolding procedure in which the states with excited electrons are integrated out.

\section{Effective electron-photon model}
\label{sec:eff_el}

The model derived in the previous section allows us to analyze the photon-conversion process, but does not provide information on how the coupling to the photons affects the charge degrees of freedom. 
We will next devise a simple few-states model which captures this back action on the electrons. 

\begin{figure}[t]
\centering
\includegraphics[width=0.9\columnwidth]{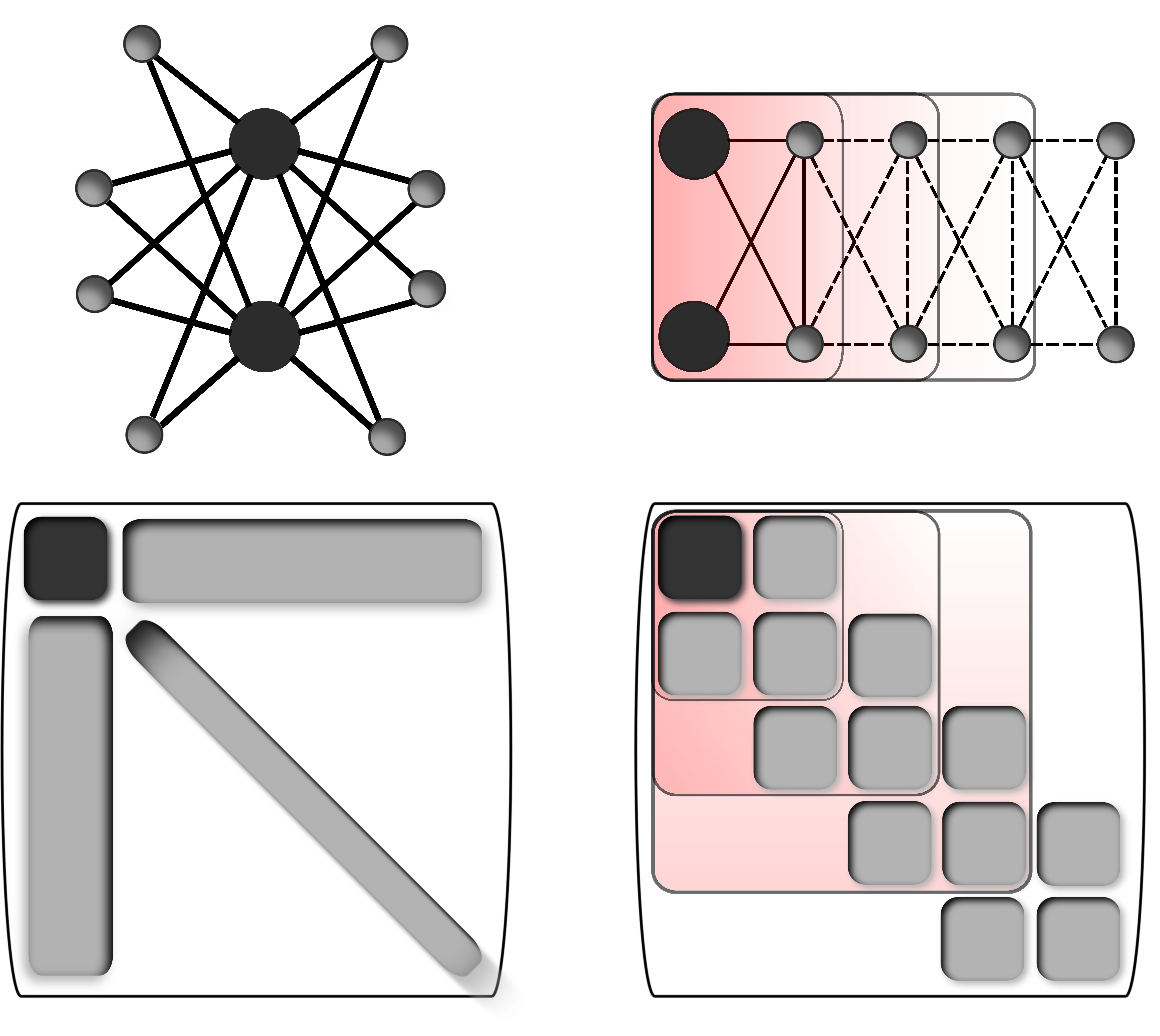}
\caption{Illustration of the mapping from a star geometry to a two-leg ladder with cross-hoppings. Block Householder transformations are successively applied to arrive at the ladder representation in the top right panel. The lower panels show a sketch of the nonzero elements of the corresponding Hamiltonian matrices. The black square represents the submatrix $\text{diag}(\Omega_2,2\Omega_1)$, which is unaffected by the transformation. 
}   
\label{fig:chain_rep}
\end{figure}

The matrix of the low-energy model
(\ref{eq:fullHam}) represents a generalized star geometry, illustrated in the top left panel of Fig.~\ref{fig:chain_rep}, in which two photon states (black dots) are coupled to $i=1,\ldots, N$ electronic states (grey dots representing 
$|v_1, \ldots, c_i, \ldots v_N\rangle |0,0\rangle$). In this geometry, it is not obvious how to optimally truncate the model and to derive a few-states model which captures the essential physics. For example, a naive truncation from $N$ to two conduction band states does not yield an appropriate model, since the effect of excitations at other $k$ points is simply ignored. Instead, we implement a procedure which is analogous to  the mapping of the Anderson impurity model from a star geometry to a chain geometry. After the latter transformation, the coupling to the first site of the chain is given by the geometric mean of all possible impurity-bath couplings.\cite{Gubernatis_2016} The first site in the chain hence represents a ``molecular bath orbital," and even after a truncation of the chain, the model captures some relevant non-local physics. 

In the following, we employ a similar mapping involving $2\times 2$ block matrices.  
In deriving this transformation, we draw on previous results for block transformations of symmetric matrices.\cite{Schreiber_1988, Rotella_1999}  
For notational convenience, we subsume some symbols in Eq.~\eqref{eq:fullHam} and write 
\begin{equation} \label{eq:fullHamSimpl}
{H}^\text{low}=
\begin{pmatrix}
	\Omega_2 & 0 & V_1 & V_2 & \dots & V_N \\
	0 & 2\Omega_1 & V_1^\prime & V_2^\prime & \dots & V_N^\prime \\
	V_1^* &(V_1^\prime)^* & \epsilon_1 & 0 & \dots \\
	V_2^* &(V_2^\prime)^* &0 & \epsilon_2 &  \\
	\vdots & \vdots & & & \ddots & 0 \\
	V_N^* &(V_N^\prime)^* & 0 &\dots & 0 & \epsilon_N 
\end{pmatrix}.	
\end{equation}	

The high degree of entanglement is evident by the off-diagonal couplings between the electronic subsystem and the photons. In analogy to the chain representation of the Anderson impurity model,\cite{Gubernatis_2016} we seek a generalized Householder transformation that can transform the problem into a tridiagonal block matrix form. Adapting the derivation of Ref.~\onlinecite{Rotella_1999} to suit the present problem of a Hermitian starting matrix, we define the following block Householder transformation
\[
U_1=\left(\begin{array}{@{}c|c@{}}
  \begin{matrix}
  1_{2\times 2}
  \end{matrix}
  & 0_{2\times N} \\
\hline
  0_{N\times 2} &
  \begin{matrix}
  h_1 
  \end{matrix}
\end{array}\right),
\]
where \footnote{ Note that $V_A$ is not a square matrix. $V_A (V_A^{\dagger}V_A)^{-1}$ is the so-called Moore-Penrose inverse of $V_A^\dagger$.}
\begin{equation} \label{eq:HVA}
	h_1
	= \boldsymbol{1} - 2V_A (V_A^{\dagger}V_A)^{-1}V_A^{\dagger}
\end{equation}
and $V_A$ is defined in terms of a still to be determined matrix $X$ as
\begin{equation} \label{eq:va}
	V_A = \begin{pmatrix}
	A_1 + X \\
	A_2 \\
	\end{pmatrix},
\end{equation}
with 
\begin{equation}
A_1 \equiv \begin{pmatrix} V_1^* & (V_1^\prime)^* \\ V_2^* & (V_2^\prime)^* \end{pmatrix}, 
\quad
	A_2 \equiv \begin{pmatrix} V_3^* & (V_3^\prime)^* \\ 
	 \vdots & \vdots \\
	   V_N^* & (V_N^\prime)^* \end{pmatrix}.
\end{equation}
(We will denote the $N\times 2$ matrix $\big($$A_1\atop A_2$$\big)$ by $A$.)
Using the invertibility of $A_1$, one can derive the following result for $X$:
\begin{equation}
	X = \sqrt{1+ \Lambda^{\dagger} \Lambda}A_1, \quad \Lambda = A_2 A_1^{-1}.
\end{equation}
To ensure that the Householder transformation changes the Hamiltonian matrix $H^\text{low}$ into the form
\begin{align}
	&U_1^\dagger H^\text{low} U_1 = \begin{pmatrix} H^\text{low}_{\text{photon}} & (h_1 A)^\dagger \\
	h_1 A & h_1 H^\text{low}_{(1,...,N),(1,...,N)} h_1 \end{pmatrix} \nonumber\\
	&\hspace{5mm}\equiv
	\left(
	\begin{tabular}{c|cl}
	$H^\text{low}_{\text{photon}}$ & $E_1^\dagger$ & $0 \hspace{3mm}\cdots$ \\
	\hline
	$E_1$ & & \\
	$0$ & & $h_1 H^\text{low}_{(1,...,N),(1,...,N)}h_1$ \\
	$\vdots$ & &  
	\end{tabular}
	\right),
	\label{eq:transformedHam}
\end{align} 
(with $H_\text{photon}^\text{low}$ denoting the first $2\times 2$ diagonal block and $H^\text{low}_{(1,...,N),(1,...,N)}$ the last $N\times N$ diagonal block) 
we demand that 
\begin{equation}
	h_1
	\begin{pmatrix} A_1 \\ A_2 \end{pmatrix} = \begin{pmatrix} E_1 \\ E_2 \end{pmatrix} 
	\equiv \begin{pmatrix} E_1 \\ 0 \end{pmatrix} .
\end{equation}
This implies
\begin{align}
	E_2 &= A_2 - 2 A_2 (V_A^\dagger V_A)^{-1}V_A^\dagger A \nonumber\\
	& = A_2 (V_A^\dagger V_A)^{-1} [ (V_A^\dagger V_A) - 2V_A^\dagger A] = 0, \label{eq:condition}
\end{align}
or $(V_A^\dagger V_A) =2V_A^\dagger A$. Writing out Eq.~\eqref{eq:condition} using Eq.~\eqref{eq:va}, we find
	$A_1^\dagger X = X^\dagger A_1$.
To solve for $X$, it is useful to define the quantity $Z = X A_1^{-1}$, which due to the above has the property\footnote{If $A_1$ is invertible, then $(A_1^{-1})^\dagger = (A_1^\dagger)^{-1}$ and from $A_1^\dagger X = X^\dagger A_1$ we have that $X^\dagger=A_1^\dagger X A_1^{-1}$. Therefore $Z^\dagger =(A_1^{-1})^\dagger A_1^\dagger X A_1^{-1} = Z$.} 
$Z^\dagger = Z$
and
\begin{equation}
	Z^2=Z^\dagger Z  = 1 + \Lambda^\dagger \Lambda
\end{equation}
with $\Lambda = A_2 A_1^{-1}$. The transformation\footnote{The second identity is a consequence of the fact that $Z^2=(Z^2)^\dagger$.} $Z^2=P D P^{-1}=P D P^\dagger$ to the diagonal matrix $D$ allows us to express $X$ as
\begin{equation}
	X = P \sqrt{D}P^{-1}A_1.
\end{equation}
In general, a diagonal non-negative $n\times n$ matrix can have at most $2^n$ square roots. To avoid any possible ambiguities we select all non-negative roots on the diagonal of $\sqrt{D}$.
 
The repeated application of such unitary 
transformations leads to the block tridiagonalization of the matrix and the ladder structure sketched in the upper right panel of Fig.~\ref{fig:chain_rep}. The second unitary transformation, $U_2$, is defined as 
\[
U_2=\left(\begin{array}{@{}c|c@{}}
  \begin{matrix}
  1_{4\times 4}
  \end{matrix}
  & 0_{4\times (N-2)} \\
\hline
  0_{(N-2)\times 4} &
  \begin{matrix}
  h_2 
  \end{matrix}
\end{array}\right),
\]
with 
$h_2$ determined by an analogous procedure as described above. 
This and subsequent transformations are designed to transform the remaining parts of the matrix to a block tri-diagonal form. The unitary matrix $U$ defined by the product 
\begin{equation}
U=U_1\cdot U_2\dots U_{N/2-2}\cdot U_{N/2-1}
\end{equation}
thus produces the desired mapping. 

Since as in the case of the Anderson impurity model, the first sites of the ladder represent molecular orbitals, the truncation of the model at this level (four-states model) still captures the coupling of the photons to the entire lattice. We will show in Sec.~\ref{sec:results} that this simple model indeed provides an accurate description 
of the photon conversion amplitudes, and a meaningful description of the back action of the photons on the electrons.  

To measure observables in the effective four-states model, we need to perform the same basis transformation on the corresponding operator matrices prior to truncation, symbolically denoted by $\mathcal{O}\rightarrow T U^\dagger \mathcal{O}U T$, where $T$ represents the truncation of the transformed operator to a given number of states. 

In particular, the photon conversion in our model will be described by the operator matrices
\begin{equation} \label{eq:SxSy}
S_x=\left(\begin{array}{@{}c|c@{}}
\sigma_x
  & 0_{2\times N} \\
\hline
  0_{N\times 2} &
  \begin{matrix}
  0_{N\times N}
  \end{matrix}
\end{array}\right) ,
\quad
S_y=\left(\begin{array}{@{}c|c@{}}
\sigma_y
  & 0_{2\times N} \\
\hline
  0_{N\times 2} &
  \begin{matrix}
  0_{N\times N}
  \end{matrix}
\end{array}\right) ,
\end{equation}
where $\sigma_i$, $i=x,y$ are the Pauli matrices. 

\begin{figure}[b] 
\centering 
\includegraphics[width=1\columnwidth]{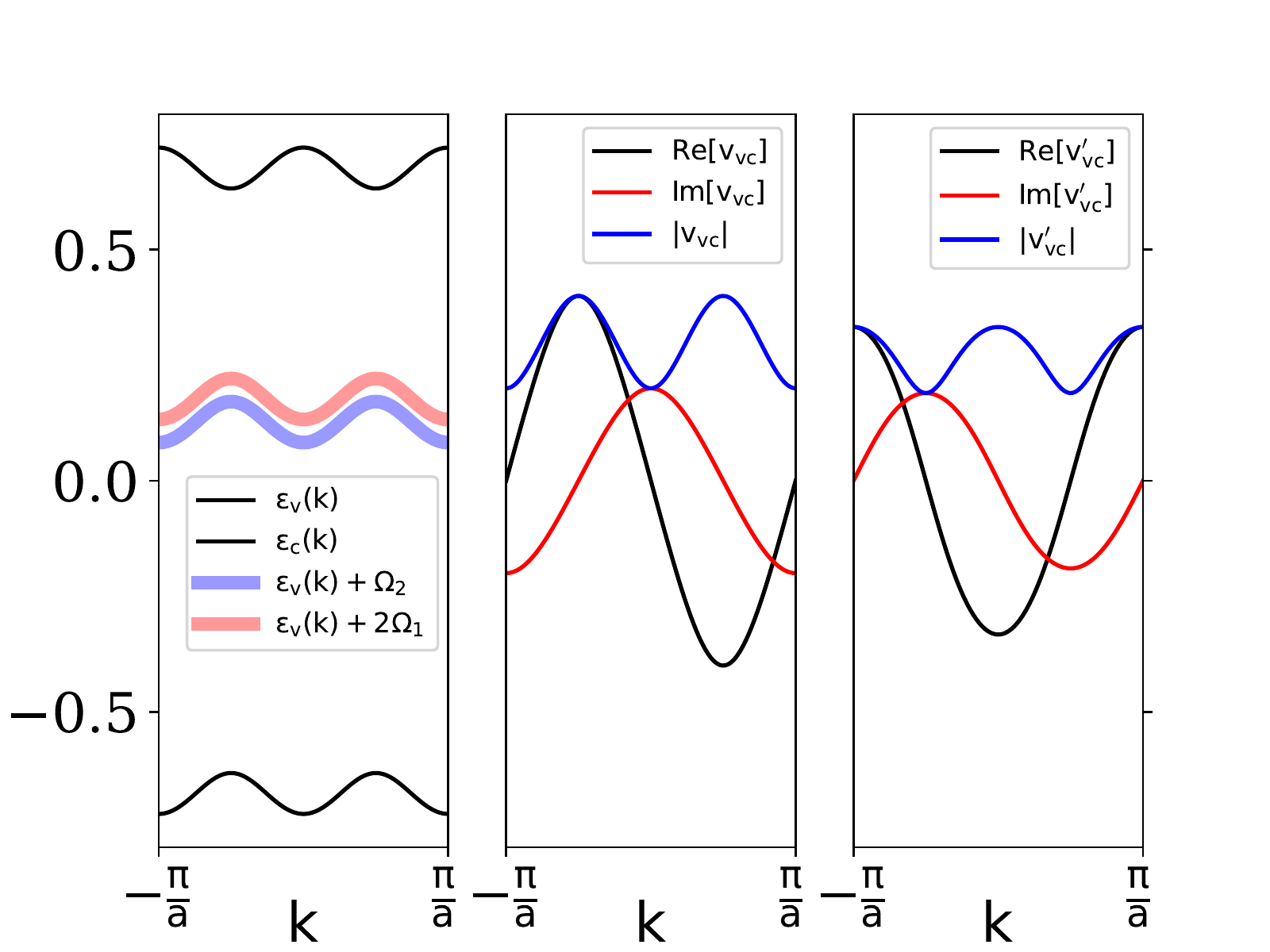}
\caption{Plot of the band structure of model (\ref{eq_model}), together with the off diagonal matrix elements of $\hat{v}(k)$ and $\hat{v}'(k)=\tfrac{\partial \hat{v}(k)}{\partial k}$, for the parameters $\Omega_2=0.853$, $t=-0.2$, $\tilde{t}=0.1$, $Q=0.6$,  
}
\label{fig:bandFig}
\end{figure}

\section{Results}
\label{sec:results}

\subsection{The model}

As in the previous sections, we consider a spinless electron system with two bands, which is coupled to two photon modes with frequencies $\Omega_1$, $\Omega_2$ and coupling constants $g_1$ and $g_2$, and we assume $\Omega_2=2\Omega_1$. 
The electronic part is given by a one-dimensional chain with a staggered potential $Q$ and bond strength  $\tilde{t}$, 
corresponding to the real-space Hamiltonian \cite{Nagaosa_2017}
\begin{equation}
	\hat{H}_0 = \sum_i 2(t - \tilde{t}(-1)^i)(\hc_{i}^\dagger \hc_{i+1} + \mathrm{h.c.}) + Q(-1)^i \hc_i^\dagger \hc_i .
\label{eq_model}
\end{equation}
This model has TRS but breaks IS if both $\tilde t$ and $Q$ are nonzero. \cite{Morimoto_2016} In a sublattice basis we can write $\hat{H}_0=\sum_k \psi^\dagger(k) h_0(k)\psi(k)$ with
\begin{equation} \label{eq:origModel}
h_0(k)=\begin{pmatrix}
-2t\cos(k) & 2i\tilde t\sin(k) + Q \\
-2i\tilde t\sin(k)+Q & 2t \cos(k) \\
\end{pmatrix}
\end{equation}
and $\psi^\dagger_k=(c^\dagger_{k}, c^\dagger_{k+\pi})$. The energy bands for the model parameters $\Omega_2=0.853$, $t=-0.2$, $\tilde t=0.1$, $Q=0.6$, as well as the corresponding matrix elements $v_{vc}$ and $v'_{vc}$ in the band basis are shown in Fig.~\ref{fig:bandFig}. In calculating $v_{vc}$ and $v'_{vc}$ we fix the arbitrary complex phase of the Bloch functions in such a manner as to keep the first entry of the eigenstates real. We will use this set-up in the following analysis.

\subsection{Tests of the few-states model}

\begin{figure}
\centering
\hspace{4.5mm}\includegraphics[width=0.9\linewidth]{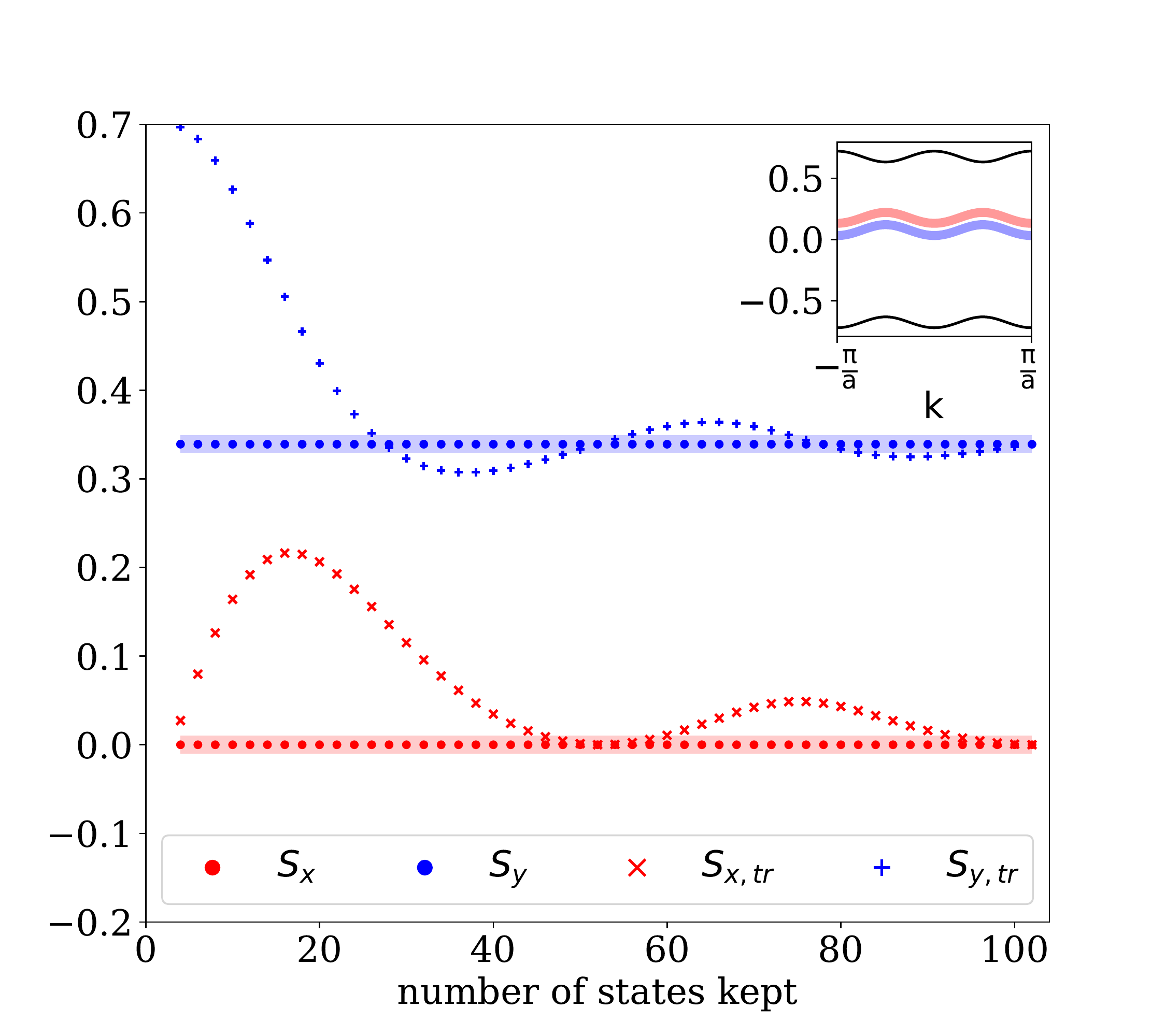}
\includegraphics[width=0.9\linewidth]{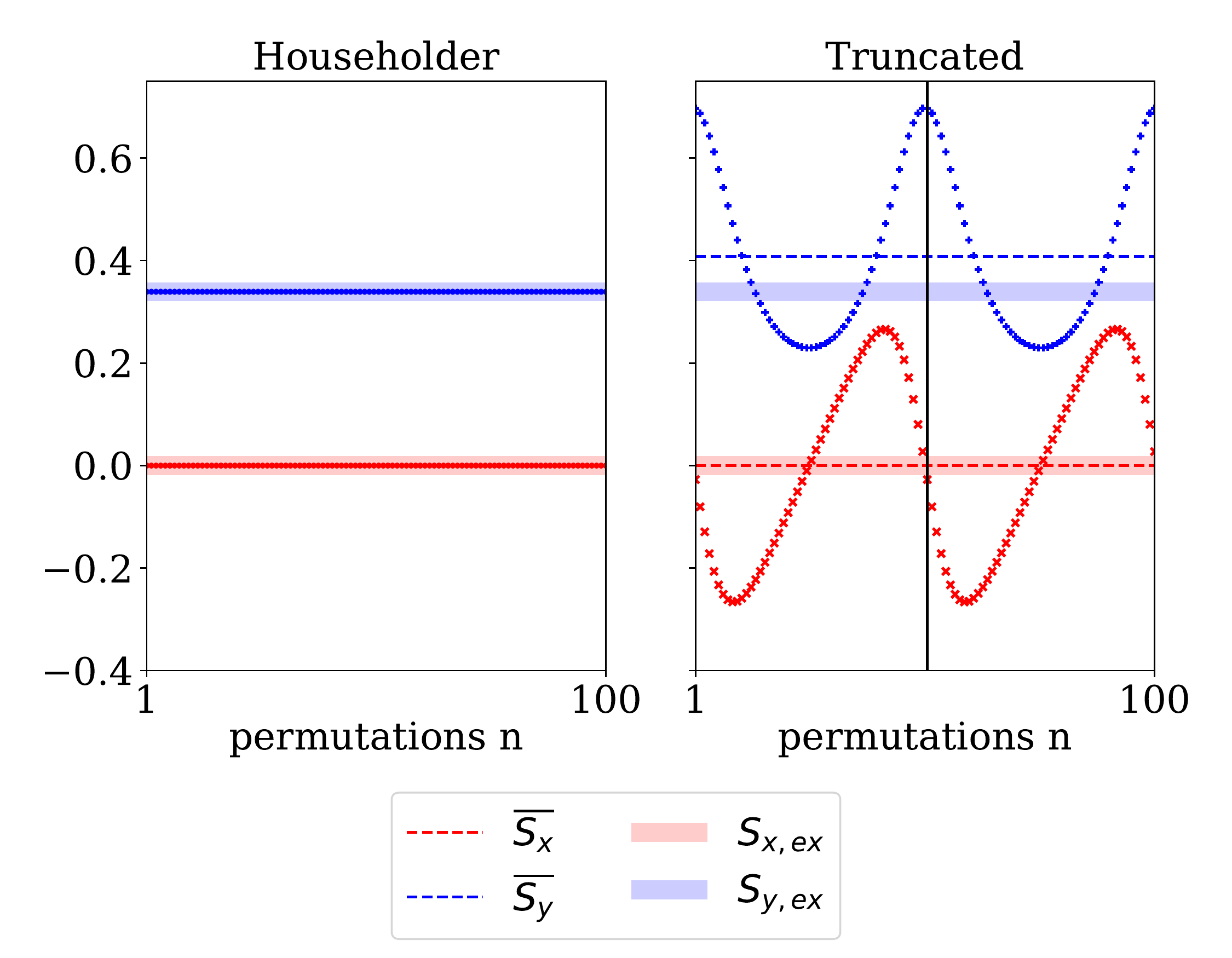}%
\caption{Test of the accuracy of the truncated models for $\langle \hat S_x\rangle$ and $\langle \hat S_y\rangle$. The light red and light blue horizontal lines show the exact reference values from the full model, while dots (crosses) indicate the values obtained by the truncated Householder model (simple truncation). The top panel shows the convergence with the number of states kept, while the bottom panels show results for different four-states models obtained by cyclic permutations of the $k$ points. Here, we consider an ordered set of $k$ points (see text). The model parameters are $\tilde{g}_1=0.035$, $\tilde{g}_2=0.005$, $\Omega_2=0.85$, $t=-0.2$, $\tilde{t}=0.1$, $Q=0.6$, $N=100$. The corresponding band structure is shown in the inset of the top panel.}
\label{fig:regularGrid}
\end{figure}

To investigate the accuracy of the few-states models derived by the Householder scheme we focus on the expectation values of $\hat S_x$ and $\hat S_y$. In the practical implementation, there is a freedom in choosing the ordering of the $k$ points in the matrix \eqref{eq:fullHamSimpl}. Let us define an ordered set of points $\tilde k_i=-\pi+(i-1)(2\pi/N)$, $i=1,\ldots, N$. The upper panel of Fig.~\ref{fig:regularGrid} shows the convergence of the truncated models towards the exact results (light red and light blue lines) with increasing number of retained states, for the choice $k_i=\tilde k_i$. The results derived by the simple truncation of the matrix (cross symbols) converge very slowly, and these models require either the full set of $k$ points, or exactly half the set of $k$ points to recover the exact reference values. On the other hand, the few-states models obtained through the Householder transformation produce the correct expectation values independent of the number of states retained, and in particular already for the smallest four-states model.  

The lower panels of Fig.~\ref{fig:regularGrid} illustrate how this result changes under cyclic permutations of the $k$ points: $\pi_\text{cycl}(\tilde k_i)=\tilde k_{i+1}$ (with periodic boundary conditions applied). The label $n$ on the horizontal axis refers to the number of cyclic permutations, i.e., the corresponding $k$ points are defined as $k_i=\pi_\text{cycl}^n(\tilde k_i)$. What is shown on the vertical axes are the expectation values of  $\hat S_x$ and $\hat S_y$ from the four-states models obtained by the Householder scheme (left panel) and by the simple truncation (right panel). Again, the light red and light blue lines indicate the exact reference values from the solution of the full model, while the dashed horizontal lines in the right panel show the average of the results over all cyclic permutations. In the simple truncation scheme, even such an average does not recover the correct result for $\hat S_y$. 

\begin{figure}[t] 
\centering 
\includegraphics[width=1.0\columnwidth]{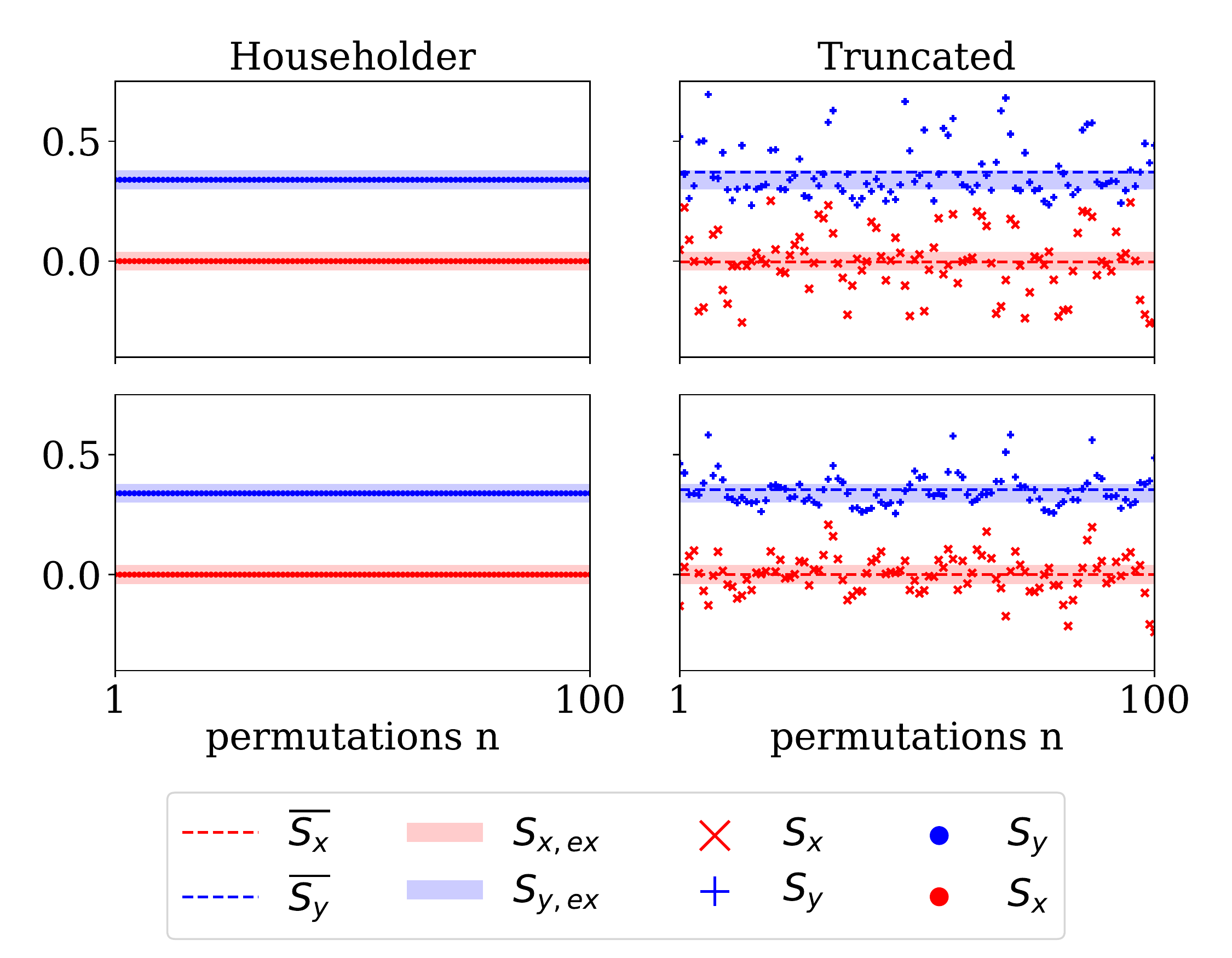}
\caption{
Results analogous to the bottom panels of Fig.~\ref{fig:regularGrid}, but for a random ordering of $k$ points (see text). The parameters are $\tilde{g}_1=0.035$, $\tilde{g}_2=0.005$, $\Omega_2=0.85$, $t=-0.2$, $\tilde{t}=0.1$, $Q=0.6$, $N=100$. 
}
\label{fig:randPerm}
\end{figure}

Next, let us consider the effect of a random permutation $\pi_\text{rand}$ of $\tilde k_i$. Figure~\ref{fig:randPerm} shows the results analogous to the bottom panels of Fig.~\ref{fig:regularGrid}, but after such a random permutation: $k_i=\pi_\text{cycl}^n(\pi_\text{rand}(\tilde k_i))$. Again, we compare the results from the four-states models obtained by the Householder scheme (left) and the simple truncation (right). The Householder approach still recovers the exact expectation values independent of the number $n$ of cyclic permutations, while the simple truncation produces a large scatter of results as a function of $n$, whose average does not reproduce the exact value for $\hat S_y$. Even though these results depend on the particular random permutation, the behavior seen in Fig.~\ref{fig:randPerm} is generic.  

These results demonstrate the usefulness of the Householder approach for deriving reliable few-states models. In fact, our numerical results suggest that in the case of $\hat S_x$ and $\hat S_y$, the few-states models derived by this scheme produce the exact expectation values independent of the number of states retained and the ordering of the $k$ points. While we cannot present a full proof, we show in Appendix~\ref{sec:shortProof} that in the case of the four-states model, once the points $k_1$ and $k_2$ have been fixed, the resulting model does not depend on the ordering of the remaining $k$ points. 
 
\subsection{Charge correlation functions}

An interesting question is how well the few-states models reproduce observables of the electronic subsystem, such as the charge correlation functions,  
\begin{equation}
	\Xi_\alpha(\omega) = \langle \hat{N}_\alpha \hat{N}_\alpha\rangle(\omega) - \langle \hat{N}_\alpha(\omega)\rangle^2, 
\end{equation}
where $\hat{N}_\alpha$ is the electronic density operator for band $\alpha=v,c$. Even though the conduction band correlation function does not scale with $N$, the measured $\Xi_\alpha$ are small. For this reason, we multiply the measurements by 100 in all figures.

We should also discuss a subtle point about the procedure used to compute the above correlation function. 
The frequency content is computed by means of the Lehmann representation, i.e., by inserting the resolution of unity in between the operators $\hat{N}_\alpha(t)$ and  $\hat{N}_\alpha(0)$. \footnote{The quantity we Fourier transform is 
$
	\Xi_c(t) = \sum_n e^{i(E_0-E_n)t} |\bra{0} \hat{N}_c\ket{n}|^2 - |\bra{0} \hat{N}_c \ket{0}|^2 
$
} It makes a difference if this is the unity resolved in the truncated or non-truncated Hilbert space. To be precise, for any operator expression, we perform the unitary transformation on all the electronic density operators in the expression and subsequently apply a truncation. 
Specifically,
\begin{equation}
	\hat{\mc{O}}_k(t) \rightarrow e^{i\hat{T}\hat{U}^\dagger \hH\hat{U}\hat{T} t} \hat{T} \hat{U}^\dagger \hat{\mc{O}}_k \hat{U} \hat{T} e^{-i\hat{T}\hat{U}^\dagger \hH \hat{U}\hat{T} t},
\end{equation}
which is consistent with the implicit rule we have employed when truncating the Hamiltonian (which is only comprised of electronic density terms). Following this convention, in the Lehmann formula, we insert the unity $1=\sum_n \ket{n}_{\text{HH}} \bra{n}_{\text{HH}}$ in between the operators $\hat{n}_{k,\alpha}(t)$ and $\hat{n}_{k,\alpha}(0)$, where $\ket{n}_{\text{HH}}$ are the eigenstates of $\hat{T}\hat{U}^\dagger \hH \hat{U}\hat{T}$ with $\hH$ the Hamiltonian operator corresponding to Eq.~\eqref{eq:fullHam}. (For an additional subtlety regarding the representation of the full model see Appendix~\ref{sec:appendixCorr}.)

\begin{figure}[t] 
\centering 
\hspace{5mm}\includegraphics[width=.9\columnwidth]{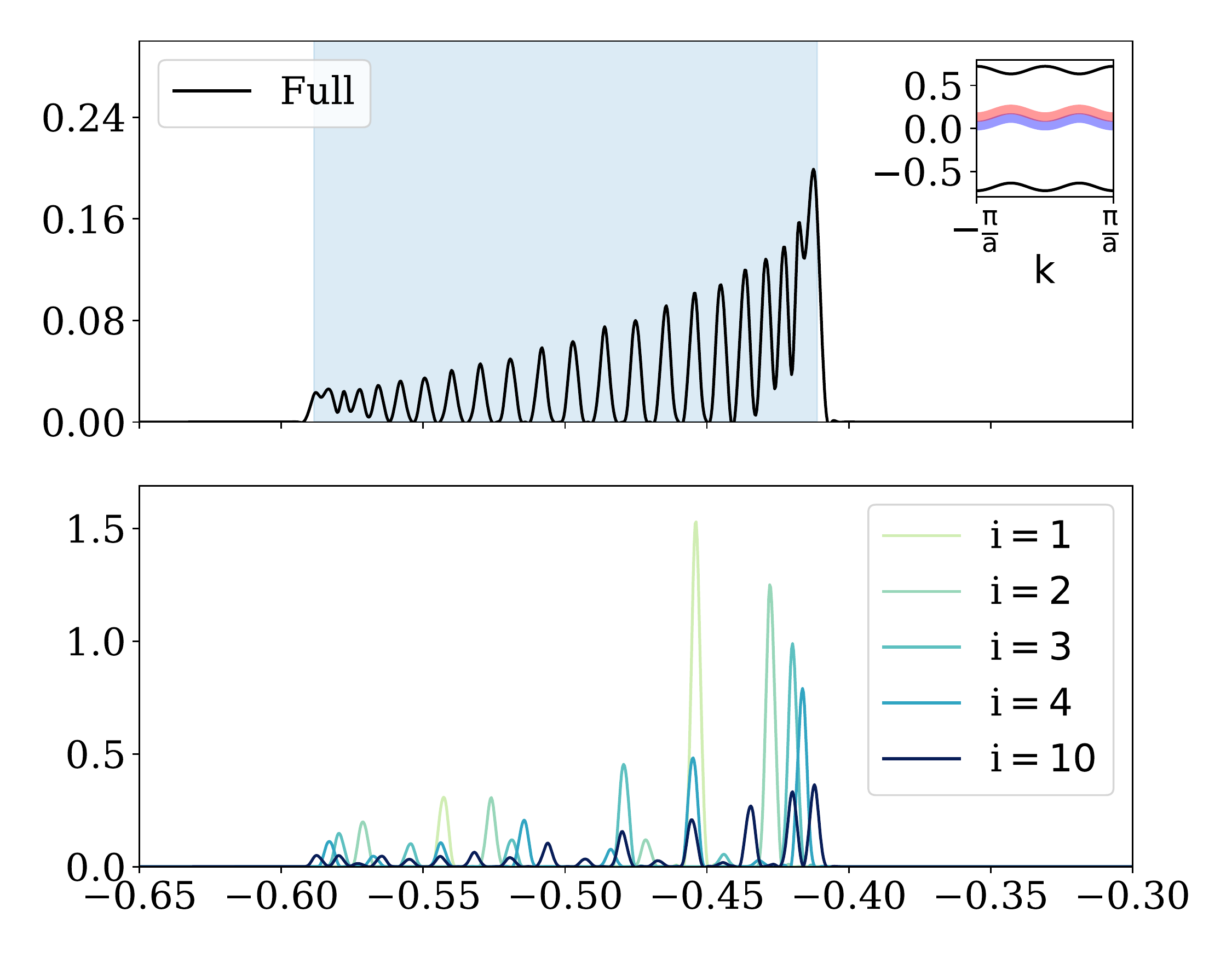}
\includegraphics[width=.85\columnwidth]{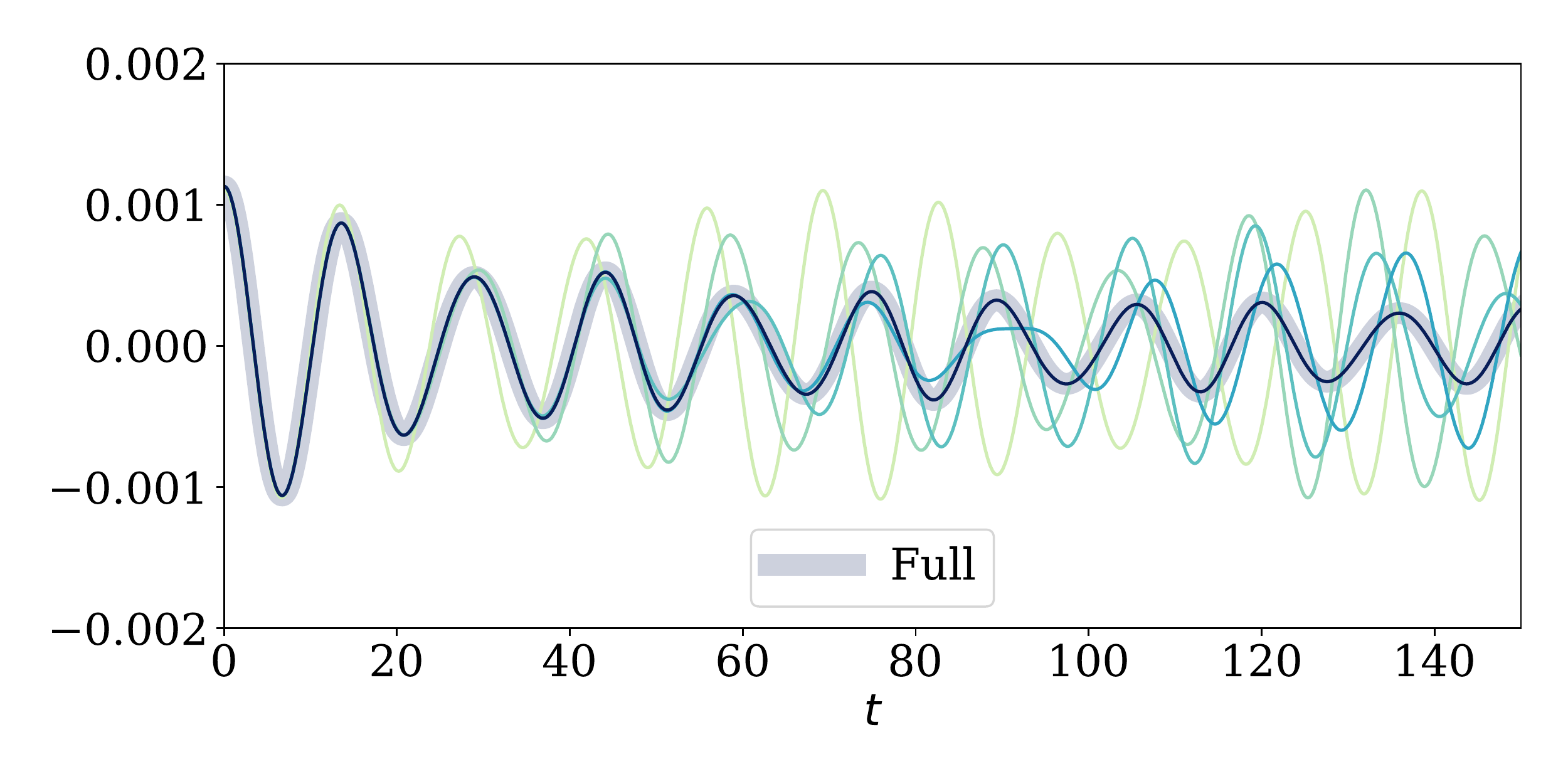}
\caption{Charge correlation function measured in the conduction band. The top panel shows the spectrum of the full model, with the shaded background indicating the energy range of possible single-particle excitations. The middle and lower panels show the results of the truncated models after $i$ Householder steps, both as a function of frequency, and as a function of time. The corresponding Hilbert space sizes are $(2i+2)$. The model parameters are $\tilde{g}_1=0.035$, $\tilde{g}_2=0.005$, $\Omega_2=0.853$, $t=-0.2$, $\tilde{t}=0.1$, $Q=0.6$, $N=100$, for which the band structure is shown in the inset. All shown results are multiplied by $100$.}
\label{fig:KSpaceCorr}
\end{figure}

In the top panel of Fig.~\ref{fig:KSpaceCorr}, we display the absolute value of the charge correlation function of the full model for $\alpha=c$. The region of nonzero weight covers the semi-continuum of energy excitations obtained from the diagonalization of model \eqref{eq:fullHam}. 
In the middle panel, results from the truncated models obtained by the indicated numbers of Householder iterations are shown. With just two ``molecular orbitals" the four-states model is of course not able to accurately capture the full range of possible excitations, but it provides two peaks which in a reasonable way represent the continuum of excitations in the full model. At the next iteration (6-states model) the energy range of possible excitations is already well captured, while after ten iterations (22-states model) the envelope of the excitation spectrum starts to be correctly captured. 
In contrast, the simple truncation produces an erratic collection of peaks that converges very slowly with increasing dimension of the truncated space (not shown). 

Alternatively, the accuracy of the Householder scheme can be assessed by looking at the time-dependent density-density correlations and in particular at the critical time after which the approximate correlation functions start to deviate from the reference data from the full model (grey line in the bottom panel of Fig.~\ref{fig:KSpaceCorr}).  The four-states model captures the first oscillation, the 6-states model the first three oscillations, while the 22-states model correctly reproduces all oscillations up to $t\approx 200$.

One may wonder if it is also possible to compute site-dependent correlation functions. 
By assuming translational invariance, we may define the charge correlation function between sites 0 and $R$ as 
\begin{equation}
	\xi_\alpha(R,t) \equiv \expval{\hat{n}_{R,\alpha}(t) \hat{n}_{0,\alpha}(0)} - \expval{\hat{n}_{R,\alpha}}\expval{\hat{n}_{0,\alpha}}.
\end{equation}
Focusing on the conduction band and using $\hc_{k,c}^\dagger = N^{-1/2}\sum_R e^{ikR} \hc_{R,c}^\dagger$, the Fourier transform $\xi_c (q,t) = \sum_R e^{-iqR} \xi_{c}(R,t)$ becomes
\begin{equation} \label{eq:xiq}
	\xi_c(q, t) = \delta_{q,0} N^{-1} \sum_{k,p} [ \expval{ \hat{n}_{k,c}(t) \hat{n}_{p,c} } - \expval{ \hat{n}_{k,c} } \expval{\hat{n}_{p,c} } ].
\end{equation}
Here, we have used that 
in our truncated space 
$\hc_{k,c}^\dagger \hc_{k',c}=\hat{n}_{k,c} \delta_{k,k'}$, see also Appendix~\ref{sec:appendixCorr}. The $\delta_{q,0}$ factor implies infinitely long-ranged correlations. For the conduction band, we find that
	$\Xi_c(t) = N \xi_c(q=0, t)$, 
which means that it is sufficient to analyze the total charge correlations. 

\subsection{Kinetic energy results}
\begin{figure}[t] 
\centering 
\includegraphics[width=0.9\columnwidth]{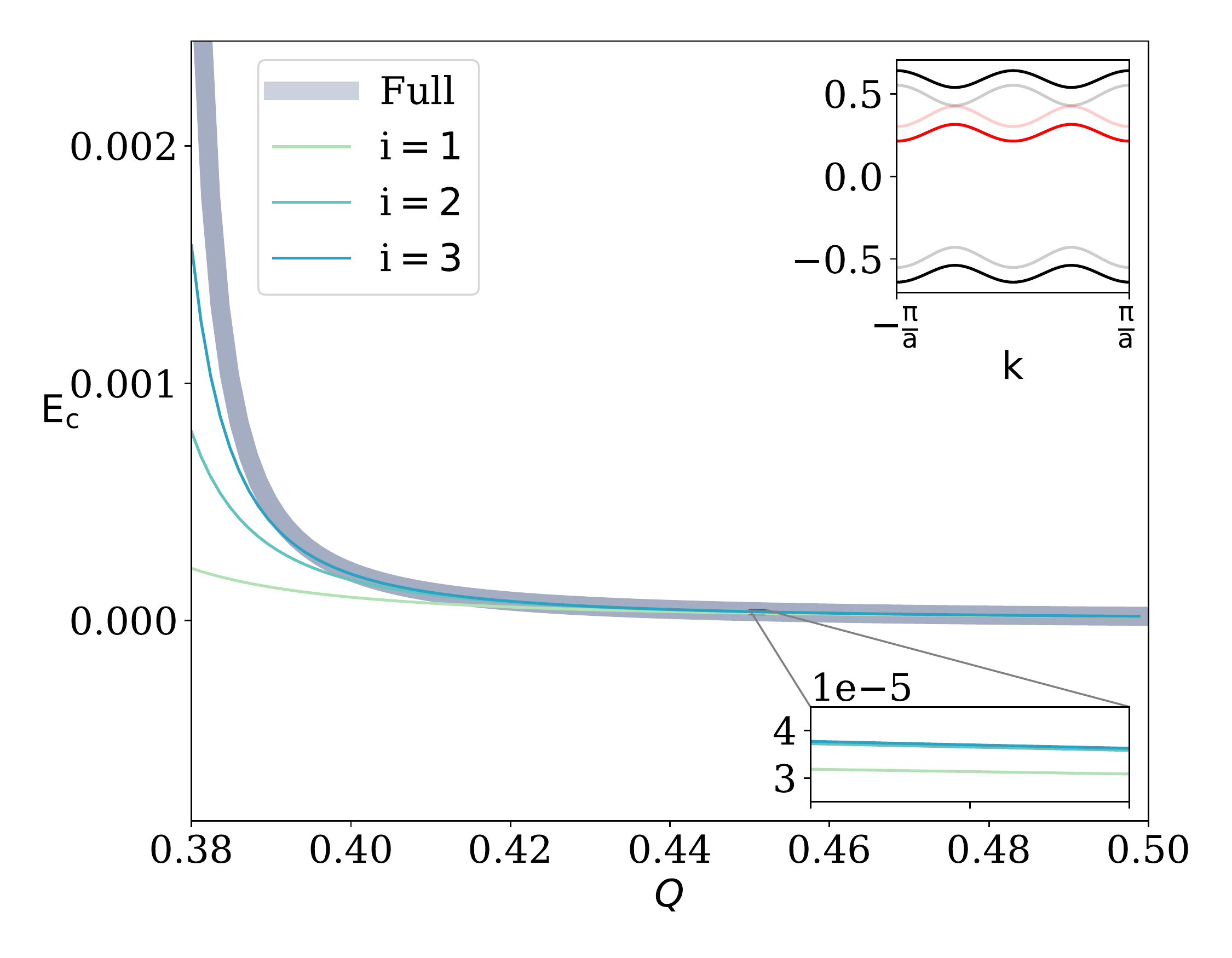}
\caption{
Kinetic energy contribution of the conduction band for different values of $Q$ in the vicinity of the band touching between the conduction band and the photon-dressed valence band (see inset: $Q=0.38$ corresponds to the semi-transparent bands and $Q=0.5$ to the opaque bands). The remaining parameters are $\tilde{g}_1=0.035$, $\tilde{g}_2=0.005$, $\Omega_2=0.85$, $t=-0.2$, $\tilde t=0.1$, $N=100$. The thick grey line shows the results of the full model, while the other lines indicate the results after $i$ Householder steps and subsequent truncation to a $(2i+2)$-states model.
}
\label{fig:kinEn1}
\end{figure}

\begin{figure}[t] 
\centering 
\includegraphics[width=1\columnwidth]{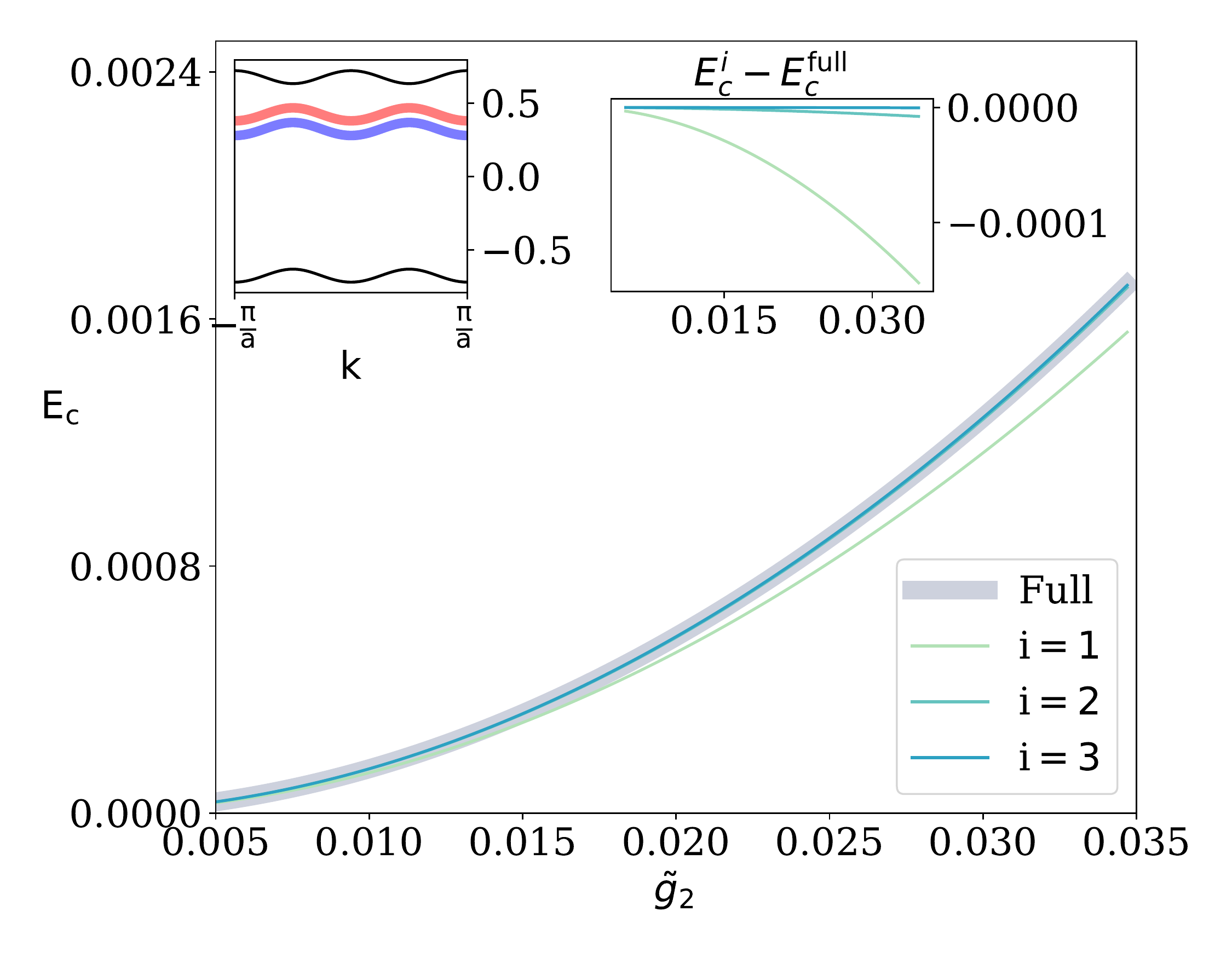}
\caption{
Kinetic energy contribution of the conduction band for different values of the photon coupling strength $\tilde{g}_2$. The parameters are $\tilde{g}_1=0.035$, $\Omega_2=0.85$, $t=-0.2$, $\tilde t=0.1$, $Q=0.6$ and $N=100$, and the corresponding band structure is shown in the left inset. The exact result is plotted as a thick grey line, while the other curves indicate the results after $i$ Householder steps and subsequent truncation to a $(2i+2)$-states model. 
The upper right inset shows the difference in $E_c$ between the full and truncated model. 
}
\label{fig:kinEn2}
\end{figure}

In Figs.~\ref{fig:kinEn1} and \ref{fig:kinEn2}, we present results for the contribution of the conduction band to the kinetic energy. Again, we compare the results of different iterations and truncations of the Householder method to the full model. 

Figure~\ref{fig:kinEn1} plots 
\begin{equation}
E_{c}=\sum_k \epsilon_c(k) \expval{\hat{n}_{k,c}},
\label{ekin}
\end{equation}
which, just like the charge correlation functions discussed previously, is a quantity that does not scale with system size in our model. The inset panel shows the modifications in the band structure as $Q$ is changed from a low value, where the photon-dressed valence band touches the conduction band (semi-transparent band structure), to the the higher values corresponding to a substantial gap (opaque band structure). The thick grey line represents the results obtained for the full model (Eq.~\eqref{eq:fullHam}). For all values of $Q$, the accuracy of the truncated Householder model increases systematically with increasing $i$, but the deviations get larger for smaller $Q$. 

Close to the value $Q=0.38$, where the bands touch, we observe a strong enhancement of the kinetic energy. In this parameter regime, electron-hole excitations become more probable due to the decreased energy gap. We caution however that even the full model must be treated with care in this region, due to the increased likelihood of multi-electron excitations to the conduction band, which are not captured in Eq.~\eqref{eq:fullHam}. Unless the coupling strengths are decreased correspondingly, even the full model description will become inaccurate. 

In general, in the presence of band crossings, new basis states must be introduced in the low-energy model, including states that have a substantial fraction of electrons in the conduction band. 

Finally, we plot in Fig.~\ref{fig:kinEn2} the kinetic energy contribution of the valence band as a function of the photon coupling strength $g_2$, for a set-up with a sufficiently large splitting between the bands that our low-energy model \eqref{eq:fullHam} is justified. The deviations between the full calculation (thick grey line) and the results from the truncated Householder models increase with increasing coupling strength, but the results of the truncated models systematically and rapidly converge towards the exact result with increasing number of states kept in the effective description (see also the right inset, which plots the difference to the exact result). These data confirm that the effective few-states models derived via the Householder transformations correctly capture the effect of the photon coupling on the electronic properties. 

\section{Conclusions}

We have introduced a minimal model describing photon conversion processes in cavity light-matter systems. If the lattice model representing the matter subsystem has TRS, but no IS, even order nonlinear processes are activated, with a transition amplitude that is related to the shift vector. This result generalizes previous analyses for lattice models driven by classical light \cite{Aversa_1995,Sipe_2000,Morimoto_2016} to the quantum domain. 
It should be noted, however, that the set-up considered in our study differs in one important respect from the Floquet study in Ref.~\onlinecite{Morimoto_2016}. There, a low-energy model was considered which describes a Floquet sideband of the valence band overlapping with the conduction band. In the present study, we considered situations with a sufficiently large gap between the photon-dressed valence band and the conduction band, and weak photon couplings, so that it is meaningful to restrict the low-energy model to states with at most one electron in the conduction band. 

While a photon-only model capturing the effect of the light-matter coupling on the photon states can be easily derived by time-dependent perturbation theory\cite{Kockum_2017} or downfolding, the long-ranged correlations induced by the photons make the derivation of an effective few-states model with electronic degrees of freedom a nontrivial task.   
We employed a generalized Householder transformation to introduce a coupling of the photons to extended molecular orbitals. This transformation enables systematic truncations to effective models with a small number of electronic degrees of freedom, which nevertheless capture the interplay between the photons and the electronic subsystem. In particular, we showed that the photon conversion processes are still accurately described even by a four-states model, while the back-action of the photons on the electronic subsystem can be at least qualitatively captured with a modest number of states. This is in stark contrast to a simple truncation in the original basis of our minimal model, which due to the highly entangled nature of the photon and electron subsystems, does not result in a meaningful description. 

We illustrated the usefulness of the Householder approach by demonstrating the fast convergence of the effective model descriptions to the full result with increasing number of states kept, for different observables related to the electronic and photonic subsystems. This is in particular the case if there exists a sufficiently large energy gap between the nonzero photon states and those with electronic excitations. As this gap shrinks, states with multiple electrons in the conduction band become more likely, and both our original model and its effective few-states descriptions will eventually break down.  

Generalizations of the Householder method to multi-mode systems or other light-matter systems with a high degree of entanglement are interesting prospects. It is also worthwhile to investigate more closely the structure of the effective orbitals generated by the Householder scheme, and to search for models in cavity QED, for which the application of the block Householder transformation results in an analytically solvable problem.  This could produce valuable insights into the nontrivial correlations induced by strong light-matter coupling.           

\acknowledgements

We thank M. Eckstein for helpful discussions. The calculations have been performed on the Beo04 and Beo05 clusters at the University of Fribourg. This work has been supported by the European Research Council through ERC Consolidator Grant No. 724103.

\appendix

\section{Gauge considerations} \label{sec:appendixGauge}

\subsection{General remarks}

Traditionally, light-matter coupled systems in the semi-classical approximation have been described in either the length gauge or velocity gauge.\cite{Taghizadeh_2017, Passos_2018} Both representations have their merits, and the effect of basis truncations in both of them has been discussed in terms of the (multi-center) PZW transformation.\cite{Schuler_2021, Golez_2019, Li_2020} Here, we demonstrate that these representations are connected by unitary transformations and hence equivalent in the non-truncated Hilbert space. 
The strategy is to pass via the length gauge, corresponding to a term $\boldsymbol{\hat r}\cdot\boldsymbol{E}$ in the Hamiltonian in first quantization. Since this breaks translational invariance (something which is also apparent in the Hamiltonian of Ref.~\onlinecite{Golez_2019}), one needs to consider the infinite volume limit. Blount\cite{Blount_1962} showed that in this limit, the matrix element of $\boldsymbol{\hat r}$ between Bloch states can be defined in terms of Eq.~\eqref{eq:BlountElement} below. In a subsequent step, we will perform yet another unitary transformation which restores translational invariance and establishes the equivalence between the Hamiltonian forms encountered in some recent literature\cite{Golez_2019, Li_2020, Dmytruk_2021} and this work. Despite the fact that these transformations are based on the infinite volume limit, our starting Hamiltonian does not break translational invariance and can therefore be studied with periodic boundary conditions. 

\subsection{Coulomb gauge Hamiltonian}

Using the notations and definitions of the main text, the Coulomb gauge Hamiltonian of the light-matter coupled system, obtained through the minimal-coupling procedure,\cite{Stokes_2020} reads 
\begin{eqnarray}
	\hat{H}_{CG} &=& \frac{1}{(2\pi)^d} \int d\bfs{k} \hat{c}_{\bfs{k},\alpha}^\dagger \bra{u_{\bfs{k},\alpha}}\hh_0(\bfs{k}-qg\OP{\bfs{A}}) \ket{u_{\bfs{k},\beta}} \hc_{\bfs{k},\beta} \nonumber\\
	&&+ \frac{\Omega_0}{2} \big( \hat{\Pi}^2 + \hA^2 \big). 
	\label{eq:HamCG}
\end{eqnarray}
For simplicity, we consider here a single photon mode with energy $\Omega_0$, $\hat{\Pi} = \tfrac{i}{\sqrt{2}}(\ha^\dagger-\ha)$, $\hA = \tfrac{1}{\sqrt{2}}(\ha+\ha^\dagger)$, and $[\hA, \hat{\Pi}]=i$, while the fermionic operators satisfy $\{ \hc_{\bfs{k},\alpha}, \hc_{\bfs{k}',\beta}^\dagger \}=(2\pi)^d \delta_{\alpha,\beta}\delta(\bfs{k}-\bfs{k}')$. Further, we define $\OP{\bfs{A}}=\hA\cdot\bfs{n}$, $\hat{\bfs{\Pi}}=\hat{\Pi}\cdot\bfs{n}$ with $\bfs{n}$ the polarization direction of the mode.

In this section, we discuss how to generate the above Hamiltonian from the one without field by means of a unitary transformation. The PZW transformation can be written as \cite{Loudon_2000}
\begin{equation} \label{eq:unitary}
	\mathcal{\hat U} = e^{igq \OP{\bfs{A}} \cdot \int d \bfs{r} \hat{\Psi}^\dagger(\bfs{r}) \bfs{r} \hat{\Psi}(\bfs{r}) } 
\end{equation}
where in the Bloch basis
\begin{align}
	&\int dr {\hat{\Psi}^\dagger}(\bfs{r}) \bfs{r} \hat{\Psi}(\bfs{r})  \\
	&= \frac{1}{(2\pi)^d}  \!\! \int \!\! d\bfs{k}d\bfs{k}' \sum_{\alpha,\beta} \hc_{\bfs{k},\alpha}^\dagger \bra{\psi_{\bfs{k},\alpha}} \hat{\bfs{r}} \ket{\psi_{\bfs{k}^\prime,\beta}} \hc_{\bfs{k}^\prime,\beta}  \nonumber\\
	&:= \frac{1}{(2\pi)^d} \!\! \int \!\! d\bfs{k} d\bfs{k}' \sum_{\alpha,\beta} \hc_{\bfs{k},\alpha}^\dagger \bfs{r}_{\alpha,\beta}(\bfs{k},\bfs{k}') \hc_{\bfs{k}^\prime,\beta} .\label{eq:rdecomp}
\end{align}
Note that this matrix element is ill defined, except in the infinite volume limit, where it is to be interpreted as \cite{Blount_1962, Ventura_2017}
\begin{align}
	&\bfs{r}_{\alpha,\beta}(\bfs{k},\bfs{k}') \nonumber\\
	&\hspace{4mm}= (2\pi)^d \delta_{\alpha,\beta} [-i\nabla_{\bfs{k}^\prime} \delta(\bfs{k}^\prime - \bfs{k}) + \xi_{\bfs{k},\alpha,\beta}\delta(\bfs{k}^\prime-\bfs{k}) ] \nonumber\\
	&\hspace{8mm} + (2\pi)^d (1-\delta_{\alpha,\beta}) \xi_{\bfs{k},\alpha,\beta} \delta(\bfs{k}^\prime-\bfs{k}),\label{eq:BlountElement}
\end{align}
which suggests to define intra- and inter-band matrix elements as follows: $\bfs{r}_{\alpha,\beta}(\bfs{k},\bfs{k}') = (\bfs{r}_i)_{\alpha,\beta}(\bfs{k},\bfs{k}') + (\bfs{r}_e)_{\alpha,\beta}(\bfs{k},\bfs{k}')$, with $(\bfs{r}_i)_{\alpha,\beta}(\bfs{k},\bfs{k}') = (2\pi)^d \delta_{\alpha,\beta} [-i\nabla_{\bfs{k}^\prime} \delta(\bfs{k}^\prime -\bfs{k}) + \xi_{\bfs{k},\alpha,\beta}\delta(\bfs{k}^\prime-\bfs{k}) ]$ and $(\bfs{r}_e)_{\alpha,\beta}(\bfs{k},\bfs{k}')= (2\pi)^d (1-\delta_{\alpha,\beta}) \xi_{\bfs{k},\alpha,\beta} \delta(\bfs{k}^\prime-\bfs{k})$ with $\xi_{\bfs{k},\alpha,\beta} = i\bra{u_{\bfs{k},\alpha}} \nabla_{\bfs{k}} \ket{u_{\bfs{k},\beta}}$. Application of the Baker-Campbell-Hausdorff (BCH) formula $e^B A e^{-B} = \sum_n \tfrac{1}{n!}[B,\ldots [B,[B,A]]]$ gives 
\begin{align}
	&\mathcal{\hat U}  \frac{1}{(2\pi)^d} \int d\bfs{k} \hat{c}_{\bfs{k},\alpha}^\dagger \bra{u_{\bfs{k},\alpha}}\hh_0(\bfs{k}) \ket{u_{\bfs{k},\alpha}} \hc_{\bfs{k},\alpha} \mathcal{\hat U}^\dagger \nonumber\\
	& \hspace{3mm}= \frac{1}{(2\pi)^d} \int d\bfs{k} \hat{c}_{\bfs{k},\alpha}^\dagger \sum_{n=0}^\infty \frac{1}{n!} \big( (-qg\OP{\bfs{A}} \cdot\nabla_{\bfs{k}})^{n} \hh_0(\bfs{k}) \big)_{\alpha,\beta} \hc_{\bfs{k},\beta} \nonumber\\
	& \hspace{3mm}=\frac{1}{(2\pi)^d} \int d\bfs{k} \hat{c}_{\bfs{k},\alpha}^\dagger \bra{u_{\bfs{k},\alpha}}\hh_0(\bfs{k}-qg\OP{\bfs{A}}) \ket{u_{\bfs{k},\beta}} \hc_{\bfs{k},\beta}. \label{eq:toProve}
\end{align}
Note that the off-diagonal elements in \eqref{eq:toProve} are generated by $\bfs{r}_e$. 
Using Eq.~(\ref{eq:toProve}), we may write the Hamiltonian in Eq.~\eqref{eq:HamCG} as 
\begin{eqnarray}
	\hH_{CG} &=& \mathcal{\hat U}  \frac{1}{(2\pi)^d} \!\! \int \!\! d\bfs{k} \hat{c}_{\bfs{k},\alpha}^\dagger \bra{u_{\bfs{k},\alpha}}\hh_0(\bfs{k}) \ket{u_{\bfs{k},\alpha}} \hc_{\bfs{k},\alpha} \mathcal{\hat U}^\dagger \nonumber\\
	&&+ \frac{\Omega_0}{2} (\hat{\Pi}^2 + \hA^2). 
\end{eqnarray}

\subsection{Connection to the dipole gauge Hamiltonian}

We define the dipole gauge Hamiltonian as 
\begin{equation}
\begin{aligned}
	&\hH_{DG} = \mc{\hat U}^\dagger \hH_{CG} \mc{\hat U}.
\end{aligned}
\end{equation}
By another application of the BCH formula, one finds 
\begin{equation}
	\mc{\hat U}^\dagger \hat{\Pi} \mc{\hat U} = \hat{\Pi} + qg \frac{1}{(2\pi)^d} \!\! \int \!\! d\bfs{k} d\bfs{k}^\prime \sum_{\alpha,\beta} \hc_{\bfs{k},\alpha}^\dagger \bfs{r}_{\alpha,\beta}(\bfs{k},\bfs{k}') \hc_{\bfs{k}^\prime,\beta},
\end{equation}	
such that 
\begin{equation} \label{eq:hamDipole}
\begin{aligned}
	&\hH_{DG} = \frac{1}{(2\pi)^d} \sum_{\alpha} \!\! \int d\bfs{k} \hat{c}_{\bfs{k},\alpha}^\dagger \bra{u_{\bfs{k},\alpha}}\hh_0(\bfs{k}) \ket{u_{\bfs{k},\alpha}} \hc_{\bfs{k},\alpha} \\
	&+\frac{\Omega_0}{2} \Big( \hat{\Pi} + qg \frac{1}{(2\pi)^d} \!\! \int d\bfs{k} d\bfs{k}^\prime \sum_{\alpha,\beta} \hc_{\bfs{k},\alpha}^\dagger \bfs{r}_{\alpha,\beta}(\bfs{k},\bfs{k}') \hc_{\bfs{k}^\prime,\beta}  \Big)^2 \\
	&+\frac{\Omega_0}{2}  \hA^2.
\end{aligned}
\end{equation}
Refs.~\onlinecite{Li_2020, Schuler_2021, Golez_2019} all used the so-called multi-center PZW transformation, motivated by the fact that the standard PZW transformation in \eqref{eq:unitary} leads to a Hamiltonian breaking translational invariance. However, $\hH_{DG}$ above is unitarily equivalent to the Hamiltonian in the mentioned references, as one can show by applying the following unitary transformation
\begin{equation}
	\hat{\tilde{\mc{U}}} = e^{-qg\OP{\bfs{A}} \frac{1}{(2\pi)^d} \int d\bfs{k} d\bfs{k}^\prime \sum_{\alpha} \hc_{\bfs{k},\alpha}^\dagger \nabla_{\bfs{k}^\prime} \delta(\bfs{k}^\prime-\bfs{k}) \hc_{\bfs{k}^\prime,\alpha} }.
\end{equation}
The action of $\hat{\tilde{\mc{U}}}$ on some operator $\hat{\mc{O}} = \frac{1}{(2\pi)^d} \int d\bfs{k} \sum_{\gamma,\delta} \hc_{\bfs{k},\gamma}^\dagger \mc{O}_{\gamma,\delta}(\bfs{k}) \hc_{\bfs{k},\delta}$ is 
\begin{equation}
	\hat{\tilde{\mc{U}}}^\dagger \hat{\mc{O}} \hat{\tilde{\mc{U}}} = \frac{1}{(2\pi)^d} \int d\bfs{k} \sum_{\gamma,\delta} \hc_{\bfs{k},\gamma}^\dagger \mc{O}_{\gamma,\delta}(\bfs{k}-qg\OP{\bfs{A}}) \hc_{\bfs{k},\delta}.
\end{equation}
The Hamiltonian in Eq.~\eqref{eq:hamDipole} thus transforms as 
\begin{align}
	&\hat{\tilde{\mc{U}}}^\dagger \hH_{DG} \hat{\tilde{\mc{U}}}\nonumber = \frac{1}{(2\pi)^d} \int d\bfs{k} \sum_\alpha \hat{c}_{\bfs{k},\alpha}^\dagger \epsilon_{\alpha}(\bfs{k}-qg\OP{\bfs{A}}) \hc_{\bfs{k},\alpha} \nonumber\\
	&+\frac{\Omega_0}{2} \hat{\tilde{\mc{U}}}^\dagger \Bigg(  \hat{\Pi} + q g\frac{1}{(2\pi)^d} \int d\bfs{k} d\bfs{k}^\prime \sum_{\alpha,\beta} \hc_{\bfs{k},\alpha}^\dagger \bfs{r}_{\alpha,\beta}(\bfs{k},\bfs{k}') \hc_{\bfs{k}^\prime,\beta}  \Bigg)^2 \hat{\tilde{\mc{U}}} \nonumber\\
	&+ \frac{\Omega_0}{2}\hA^2. 
	\label{eq:hamDipoleTilde}
\end{align}
Using
$$
\hat{\tilde{\mc{U}}}^\dagger \hat{\Pi} \hat{\tilde{\mc{U}}} = \hat{\Pi} + iqg \frac{1}{(2\pi)^d} \int d\bfs{k} d\bfs{k}^\prime \sum_{\alpha,\beta} \hc_{\bfs{k},\alpha}^\dagger \nabla_{\bfs{k}^\prime}(\bfs{k}^\prime-\bfs{k}) \hc_{\bfs{k}^\prime,\beta}
$$
and 
$$
\begin{aligned}
&\hat{\tilde{\mc{U}}}^\dagger  \Bigg(qg \frac{1}{(2\pi)^d}\int d\bfs{k} d\bfs{k}^\prime \sum_{\alpha,\beta} \hc_{\bfs{k},\alpha}^\dagger \bfs{r}_{\alpha,\beta}(\bfs{k},\bfs{k}') \hc_{\bfs{k}^\prime,\beta} \Bigg) \hat{\tilde{\mc{U}}} \\
&\hspace{10mm}= -iqg \frac{1}{(2\pi)^d} \int d\bfs{k} d\bfs{k}^\prime \sum_{\alpha,\beta} \hc_{\bfs{k},\alpha}^\dagger \nabla_{\bfs{k}^\prime}(\bfs{k}^\prime-\bfs{k}) \hc_{\bfs{k}^\prime,\beta} \\
&\hspace{14mm}+ qg \frac{1}{(2\pi)^d} \int d\bfs{k} \sum_{\alpha,\beta}\hc_{\bfs{k},\alpha}^\dagger \xi_{\bfs{k}-qg\OP{\bfs{A}}, \alpha,\beta}\hc_{\bfs{k},\beta} \\
\end{aligned}
$$	
we get
\begin{align}
	&\hat{\tilde{\mc{U}}}^\dagger \hH_{DG} \hat{\tilde{\mc{U}}} = \frac{1}{(2\pi)^d} \int d\bfs{k} \sum_\alpha \hat{c}_{\bfs{k},\alpha}^\dagger \epsilon_{\alpha}(\bfs{k}-qg\OP{\bfs{A}}) \hc_{\bfs{k},\alpha} \nonumber\\
	&\hspace{3mm}+\frac{\Omega_0}{2} \Bigg( \hat{\Pi}  + qg \frac{1}{(2\pi)^d}\int d\bfs{k} \sum_{\alpha,\beta} \hc_{\bfs{k},\alpha}^\dagger \xi_{\bfs{k}-qg\OP{\bfs{A}}, \alpha,\beta} \hc_{\bfs{k}^\prime,\beta}\Bigg)^2  \nonumber\\
	&\hspace{3mm}+\frac{\Omega_0}{2}\hA^2, \label{eq:hamDipoleTildeFinal}
\end{align}
which corresponds to an infinite volume version of the dipolar Hamiltonian in momentum space [Eq.~(54)] presented in Ref.~\onlinecite{Li_2020}.

One could in principle also compute the results of the main text in one of these alternative gauges. However, the observable related to photon conversion (Eq.~\eqref{eq:SxSy}) can be expected to become a mixture of photon and electron operators if one applies the transformations above. Hence, the analysis relating photon conversion to the shift vector is most conveniently done in the Coulomb gauge.

\section{Downfolding}
\label{app_downfolding}
In the downfolding approach \cite{Rychkov_2015}, we split a generic Hamiltonian into low-energy and high-energy subspaces, 
\begin{equation}
	H = \begin{pmatrix} H_{ll} & H_{lh} \\ 
	H_{hl} & H_{hh} \end{pmatrix}.
\end{equation}
In the present context, $H_{ll}$ corresponds to the top left $2 \times 2$ block of Eq.~\eqref{eq:fullHam} or Eq.~\eqref{eq:fullHamSimpl}. To obtain an effective model for the low-energy space 
we can iteratively solve the 
eigenvalue equation 
$H_{\textrm{eff}}(\epsilon) |\psi_l\rangle = \epsilon |\psi_l\rangle$ with 
\begin{equation}
	H_{\textrm{eff}}(\epsilon) \equiv H_{ll} + H_{lh} \frac{1}{\epsilon-H_{hh}}H_{hl}.
\end{equation}
To 
a first approximation one may use the eigenenergy of the unperturbed low-energy space, $\epsilon=\Omega_2=2\Omega_1$. 
This yields   
\begin{equation}
	H_{\textrm{eff}}(\Omega_2) = \begin{pmatrix} \Omega_2 + d_z & d_{\mathrm{down}} \\
	d_{\mathrm{down}}^* & 2\Omega_1-d_z \end{pmatrix}
\end{equation}
with
\begin{eqnarray}
	d_{\mathrm{down}} &=& \sum_j \frac{V_j V_j^{\prime *}}{\Omega_2 - (\epsilon_c(k_j)- \epsilon_v(k_j))},\\
	d_z 
	&=& \frac{1}{2}\sum_j \frac{ \abs{V_j}^2 - \abs{V_j^\prime}^2 }{2\Omega_1 - (\epsilon_c(k_j) - \epsilon_v(k_j))}.
\end{eqnarray}  
This $H_\text{eff}$ is identical to the effective Hamiltonian obtained from the second order perturbation calculation.

\section{Permutation of $k$ points} 
\label{sec:shortProof}
The numerical evidence presented in the main text suggests that the four-states models derived via the Householder scheme do not depend on the ordering of the $k$ points in the original Hamiltonian matrix. 
Here, we will provide a proof that for given $k_1$ and $k_2$ (states selected in the first Householder step), the result does not change upon permutation of the remaining $k$ points. 

We have
\begin{equation}
	A_2 \rightarrow \begin{pmatrix} V_{\pi(3)} & V_{\pi(3)}^\prime \\
	V_{\pi(4)} & V_{\pi(4)}^\prime \\
	\vdots \\
	V_{\pi(N)} & V_{\pi(N)}^\prime \\ \end{pmatrix}
	:=UA_2,
\end{equation}
where $\pi(m)$ denotes a permutation of the index $m$. Due to the fact that permutation matrices are orthogonal (and hence unitary since they have real entries), we have $V_A^{\prime \dagger} V_A^\prime =V_A^{\dagger} V_A $ and Eq.~\eqref{eq:HVA} becomes $h_1^\prime= \boldsymbol{1} - 2V_A^\prime (V_A^{\dagger}V_A)^{-1}V_A^{\prime \dagger}$. Equation~\eqref{eq:transformedHam} now takes the form 
\begin{align}
& h_1^\prime H^\text{low}_{(1,...,N),(1,...,N)} h_1^\prime = \left(\begin{array}{@{}c|c@{}}
  \begin{matrix}
  1_{2\times 2}
  \end{matrix}
  & 0_{2\times (N-2)} \\
\hline
  0_{(N-2)\times 2} &
  \begin{matrix}
  U 
  \end{matrix}
\end{array}\right) \nonumber\\
&\hspace{10mm}\times h_1 H^\text{low}_{(1,...,N),(1,...,N)} h_1 
\left(\begin{array}{@{}c|c@{}}
  \begin{matrix}
  1_{2\times 2}
  \end{matrix}
  & 0_{2\times (N-2)} \\
\hline
  0_{(N-2)\times 2} &
  \begin{matrix}
  U^\dagger 
  \end{matrix}
\end{array}\right),
\end{align} 
leading us to conclude that $H^\text{low}_{(1,2),(1,2)}$ is unaffected by the permutations of $k_3$ through $k_N$. 

\section{Details on the charge correlation functions} \label{sec:appendixCorr}
In this appendix, we provide further details on how the form of Eq.~\eqref{eq:xiq} was obtained. We will consider the string of operators 
$$
	e^{i\hH t}\hc_{k_l, c}^\dagger \hc_{k_m, c} e^{-i\hH t} \hc_{k_i, c}^\dagger \hc_{k_j, c},
$$
which will be evaluated in our truncated space. Let $\mc{B}$ denote the Hilbert space comprised of all the states entering into Eq.~\eqref{eq:fullHam}. Denoting an arbitrary state in $\mc{B}$ as $\ket{B}$, we have that
$$
	\hc_{k_i, c}^\dagger \hc_{k_j, c} \ket{B} = \ket{ v_{k_1}... \{ \tiny c_{k_i}  v_{k_i} \} ... 0 ...v_{k_N} 
	} \langle v_{k_1} ...c_{k_j} ...v_{k_N} | B \rangle
$$
if $i\neq j$, and effectively that $\bra{B'}\hc_{k_i,c}^\dagger \hc_{k_j,c} \ket{B} = \bra{B'}\hat{n}_{k_i,c} \ket{B} \delta_{i,j}$ if $\ket{B'}$ is another state in $\mc{B}$. To obtain frequency dependent quantities, we will employ the Lehmann representation and insert unity in the basis $\mc{B}$: 
\begin{equation}
	\sum_n \bra{B'} e^{i\hH t}\hc_{k_l,c}^\dagger \hc_{k_m,c} e^{-i\hH t} \ket{n}\bra{n} \hc_{k_i,c}^\dagger \hc_{k_j,c} \ket{B}.
\end{equation}
Given that $\ket{n}\in \mc{B}$ together with the fact that $e^{i\hH t}$ with $\hH$ in Eq.~\eqref{eq:fullHam} induces transitions within $\mc{B}$, the following simplification will arise 
\begin{align}
	&\sum_{l,m,i,j} \sum_n \bra{B'} e^{i\hH t}\hc_{k_l,c}^\dagger \hc_{k_m,c} e^{-i\hH t} \ket{n}\bra{n} \hc_{k_i,c}^\dagger \hc_{k_j,c} \ket{B} \nonumber\\
	&=\sum_{l,i} \sum_n \bra{B'} e^{i\hH t}\hat{n}_{k_l,c} e^{-i\hH t} \ket{n}\bra{n} \hat{n}_{k_i,c} \ket{B} \nonumber\\
	&=\sum_n \bra{B'} e^{i\hH t} \hat{N}_c e^{-i\hH t} \ket{n}\bra{n} \hat{N}_c \ket{B} .
\end{align}
It is important to stress that inserting unity in the way done above involves an implicit truncation to $\mc{B}$.

\bibliographystyle{apsrev4-1}
\bibliography{bibliography.bib}

\end{document}